\RequirePackage{fix-cm}
\documentclass[twocolumn]{svjour3}
\smartqed
\usepackage{amsmath,amssymb,amsbsy,amsfonts,amsopn,amstext,amsxtra,euscript,amscd,graphicx,subfigure}
\usepackage{textcomp,enumerate,float,bm,setspace,boxedminipage,color,rotating}
\usepackage[english]{babel}
\usepackage{multirow}
\usepackage[table]{xcolor}
\usepackage{threeparttable}
\usepackage[colorlinks,linkcolor=blue,anchorcolor=blue,citecolor=blue]{hyperref}

\begin{document}
\title{Multi-Matrix Verifiable Computation}

\author{Yan He \and
        Liang Feng Zhang 
}

\institute{Yan He \at
              Chinese Academy of  Sciences, Shanghai Institute of  Microsyst $ \& $ Information Technology, Shanghai 200050, P.R. China \\
              ShanghaiTech University, School of Information Science $ \& $ Technology, Shanghai 201210, P.R. China\\
              University of Chinese Academy of Sciences, P.R. China\\
              \email{heyan@shanghaitech.edu.cn} 
           \and
           Liang Feng Zhang \at
              ShanghaiTech University, School of Information Science $ \& $ Technology, Shanghai 201210, P.R. China\\
              \email{zhanglf@shanghaitech.edu.cn}
}

\date{Received: date / Accepted: date}

\maketitle

\begin{abstract}
The problem of securely  outsourcing  computation to cloud servers has attracted a large amount of  attention in
 recent years.
 The verifiable computation of
Gennaro, Gentry, Parno (Crypto'10)  allows a client to verify the server's
computation  of a function with  substantially less time
 than  performing the outsourced  computation from scratch.
 In a multi-function model (Parno, Raykova, Vaikuntanathan; TCC'12)  of verifiable computation, the process
 of encoding function and the process of preparing input are decoupled such that
 any client can freely submit a computation request on its input, without
 having to generate an encoding of the function in advance.
In this paper, we propose a multi-matrix verifiable computation  scheme that allows the secure outsourcing
of the matrix functions over a finite field.
Our scheme is outsourceable. When it is used to outsource
$m$ linear functions, the scheme is roughly $m$ times faster and has less communication cost than the previously
best known  scheme  by Fiore and Gennaro (CCS'12), both in the client-side computation and in the server-side computation.
We also show the  cost saving  with detailed implementations.
\end{abstract}
\section{Introduction}
Cloud computing \cite{MG11,ACB10,CEM12} allows the resource-restricted clients to outsource the storage of their data and heavy  computations on the data  to the powerful cloud servers in a pay-per-use manner, which is both
  scalable and economical.
The outsourcing paradigm  however incurs  many security concerns \cite{GGP10} such as
 how to
 ensure  the outsourced   computations will be done correctly.
The powerful cloud servers are not fully  trusted and
  may  have strong financial incentives \cite{PRV12} to
  run   extremely fast
but incorrect computations, in order to free up the valuable computing time  or even benefit from
providing  incorrect  results.
 Outsourcing
 computation is useful only when the servers' results  are reliable.

 The problem of  securely outsourcing computations to clouds has
been  intensively studied  in recent years.
Numerous  solutions \cite{GGP10,GW13,BCCT12} have been proposed and optimized   for many different  scenarios.
Among them is the verifiable computation    of Gennro, Gentry and Parno \cite{GGP10}, which allows the
client to outsource the computation of a function $f$  as follows: first of all, the client runs an expensive
but one-time computation to produce an encoding of $f$ to the cloud server;
afterwards  in order to outsource the work of computing $f(x)$ for any input $x$, the client performs an efficient
computation to prepare an encoding of the input $x$ to the server; given two encodings, the server returns both  $y=f(x)$
and a cryptographic proof for its work; and finally the client efficiently verifies the
server's result with the proof.  The cryptographic proof is designed such that no
malicious server is able to persuade the client to accept any incorrect results.
The process of input preparation  and result verification should be substantially faster than computing
$f(x)$ from scratch.
The one-time effort of encoding $f$ can be amortized over the
computation  of $f$ on multiple inputs, which gives an amortized model for verifiable computation.

Following Gennaro, Gentry and Parno \cite{GGP10} there is a long line of works
 that enable  the secure outsourcing of both
  functions \cite{CKV10,AIK10,CKLR11} as generic as any  boolean circuits and
the specific functions such as polynomials   and matrices
\cite{BGV11,FG12,CFGV13}.
In all of these schemes,
  the process of preparing
$x$ heavily depends on the protocol parameters that are generated in the early process of encoding
$f$. This dependency not only requires the client to put in a large initial computational investment
before actually being able to prepare an input $x$ for delegation, but also
 requires the client to prepare the same input $x$ multiple times, whenever
 the computation of different  functions on the same input $x$ is to be delegated.
As a result, the dependency incurs significant latency
in the client-side computations.

In order to lift the dependency of input preparation on function-related protocol parameters,   Parno, Raykova, and  Vaikuntanathan \cite{PRV12}
introduced  the  multi-function model for verifiable computation  where the
process of encoding $f$ is decoupled from the process of preparing $x$ such that
any input can be preprocessed before the     functions to be outsourced  are   actually known.
In particular, they constructed a multi-function verifiable computation scheme using key-policy attribute-based encryption \cite{GPSW06,SW05} that has
outsourced decryption.  Their scheme allows the delegation of all functions that can be
covered by the permissible policies of the underlying attribute-based encryption scheme.
More precisely, this is a family of functions that can be converted into  polynomial-size boolean formulas.
While converting any function into a boolean formula is feasible in theory, doing so in practice
may incur significant loss of efficiency \cite{NP06} and  the resulting protocol would be
 prohibitively expensive.

Fiore and Gennaro \cite{FG12} initiated a study of really efficient multi-function verifiable computation schemes
for  specific classes of functions. Based on the homomorphic weak pseudorandom functions,
they constructed a  scheme for linear functions, which have a large quantity of
applications in scientific and engineering computations
{\color{black} \cite{CHL15,LLH13,LLH14,LLH15,Geo12,Mey001,TMW10,ZZW11,Gib97,MRA12}}
  as a special subset of the matrix functions.
  In particular,  for outsourcing linear function computations, their scheme is faster than
\cite{PRV12} by a logarithmic (in the size of the underlying finite field) factor in both the client-side computation and the
server-side computation.

A verifiable computation scheme is said to be outsourceable if the client-side computation
  for input preparation and result verification  is
substantially faster than   computing $f(x)$ from scratch.
 While the scheme of \cite{FG12} is much faster than \cite{PRV12},
it is not outsourceable when only one function is to be outsourced.
This is different from most of the previous works such as \cite{BGV11,FG12,CFGV13}.
In particular, when we consider the delegation of multiple matrices, one has to invoke the
scheme of \cite{FG12} multiple times, where the number of invocations is
equal to the total number of rows in these matrices.
In most applications  the dimension of the
matrices is huge. This would cause unnecessary repetitions and results in  unnecessary consumption of the client's precious computing resources.
\subsection{Our Contributions}
In this paper, we
propose a multi-matrix verifiable computation scheme where the
outsourced family of function consists of all $m\times d $ matrices over a finite field
$\mathbb{Z}_p$, where $m,d>0$ are integers and $p$ is  a   prime.
To the best of our knowledge, this is the first multi-function verifiable computation scheme for
matrix functions.
By interpreting the rows of any $m\times d$ matrix as $m$ linear functions,
our scheme enables   the delegation and
verification of $m$ linear functions in every execution.
Our scheme is outsourceable in the sense that even if it is used to delegate only one matrix function
the client can still benefit from a verification that is substantially faster than
performing the matrix-vector multiplication from scratch.
When the  scheme is used to delegate $m$ linear functions, it  outperforms  the construction
of \cite{FG12} by a factor of $m$,  both in the client-side computation and in the server-side
computation. We implemented both schemes. Our implementation shows that
our multi-matrix verifiable computation scheme is roughly $m$ times faster.
\subsection{Techniques}
Fiore and Gennaro \cite{FG12} constructed  a multi-function verifiable computation scheme for the family
${\cal F}=\mathbb{Z}_p^d$ of linear functions over a finite  field
$\mathbb{Z}_p$, where $d>0$ is an integer and $p$ is a   prime.
Their scheme  uses a cyclic group $\mathbb{G}=\langle g\rangle$
of order $p$ which is generated by $g$.
The scheme chooses $d$ group elements
$R_1,\dots,R_d\leftarrow \mathbb{G}$ as public parameters.
The preprocessing of any function
${\bf f}=(f_1,\ldots,f_d)\in {\cal F}$ is done by
 computing
a tag $W_j=g^{\alpha f_j}\cdot R_j^k$ for every $j\in[d]$, where $k,\alpha\leftarrow \mathbb{Z}_p$ are randomly chosen integers modulo $p$.
The preparation of any input ${\bf x}=(x_1,\ldots,x_d)\in \mathbb{Z}_p^d$
is done by  computing a key $VK_{\bf x}=\prod_{i=1}^d R_j^{x_j}$ for future verification.
Given the encoding $({\bf f}, W_1,\ldots,W_d)$  of the function $\bf f$ and the input
$\bf x$, the server
computes and returns both the result $y=\sum_{j=1}^d f_j x_j$ and a cryptographic proof  $V=\prod_{j=1}^d W_j^{x_j}$.
The client-side verification is done by checking the equality
$V=g^{\alpha y}\cdot  (VK_{\bf x})^k$.
It was shown that no polynomial-time server is able to persuade the client
to accept a result $\hat{y}\neq y$ with a proof $\hat{V}$, assuming that
the DDH problem is hard in $\mathbb{G}$.
Comparing with \cite{PRV12},
their scheme
results in at least   logarithmic
 speed-up    in both the client-side computation and the server-side computation.
 It is a multi-function scheme as the process of encoding $f$ is completely decoupled from
 that of preparing $\bf x$.

 In this paper we consider the more general setting of outsourcing
the family ${\cal F}_{m,d}=\mathbb{Z}_p^{m\times d}$
of $m\times d$ matrix functions over the finite field $\mathbb{Z}_p$.
We interpret any matrix ${\bf F}\in {\cal F}_{m,d}$ as a function that takes any (column) vector
${\bf x}\in \mathbb{Z}_p^d$ as input and outputs
${\bf y}={\bf F}{\bf x}$.
We note that the scheme of \cite{FG12} can be invoked
multiple times to deal with every row of the  matrix
$\bf F$ as a linear function. However, that will incur significant loss of efficiency at the client-side
as long as $m$ is large.
Our idea of delegating matrix functions is simple. On one hand, we observe that any  matrix function
 $\bf F$ can be considered as
 a set of $m$ linear functions
 $F_1=(F_{1,1},\ldots, F_{1,d}),\ldots,
 F_m=(F_{m,1}, \ldots, F_{m,d})$
 and for
  any input  ${\bf x}=(x_1,\ldots,x_d)^\top\in \mathbb{Z}_p^d$,
  the computation of  ${\bf  y}={\bf Fx}$ can be considered as
  a set of $m$ linear function evaluations:
  $y_1=\sum_{j=1}^d F_{1,j} \cdot x_j,\ldots,
  y_m=\sum_{j=1}^d F_{m,j}\cdot x_j$.
  On the other hand, we observe that
  if the $m$ linear functions can be somehow
  combined as one linear function and the
  $m$ results from the cloud server can be similarly combined
  and then verified in the vein of \cite{FG12},   the client-side work
  will be accelerated by a factor of around $m$, which can be an
   essential cost saving  as long as
  $m$ is large.
 A canonical way of combining all rows of $\bf F$ is done by computing
 their linear combinations.
 Let ${\bf r}=(r_1,\ldots,r_m)\in \mathbb{Z}_p^m$ be randomly chosen.
Then the combined function will be
${\bf s}=(s_1,\ldots,s_d)={\bf r}{\bf F}$.
In order to employ the  scheme of \cite{FG12}, the client in our scheme
computes a tag $W_j=g^{s_j}\cdot R_j^k$  for every $j\in[d]$ and then gives
both $\bf F$ and $W=(W_1,\ldots, W_d)$ to the cloud server.
In order to delegate the computation of ${\bf Fx}$, the client generates
$VK_{\bf x}=\prod_{j=1}^d W_j^{x_j}$ for future verification and simply gives
$\bf x$ to the server.
The server computes and returns both the result ${\bf y}={\bf Fx}$
and a proof $V=\prod_{j=1}^d W_j^{x_j}$.
In the verification, the client could have to check the equality
$V=g^{{\bf sx}}\cdot (VK_{\bf x})^k$.
Our method of combining linear functions was chosen such that
${\bf sx}={\bf r}{\bf F} {\bf x}={\bf ry}$, due to the associative law of matrix multiplications.  As a consequence, the
verification can be done by checking
the equality $V=g^{\bf ry} \cdot (VK_{\bf x})^k$. And in order to do so, the client only
needs to keep
$(k, {\bf r})$ as a private verification key, which is associated with the
specific function $\bf F$. In the text we show that no cloud server can persuade
 the client to accept a wrong result
$\hat{\bf y}\neq {\bf y}$ with an altered proof
$\hat{V}$, except with negligible probability.
The scheme of \cite{FG12} can be considered as an instantiation of
our multi-matrix verifiable computation scheme with $m=1$. The technique of
combining all functions as a single one to speed-up verification
may have independent interest.

\subsection
{Efficiency Analysis}
Our multi-matrix verifiable computation achieves amortized efficiency in delegating and verifying several matrices ${\bf F}_1,\dots,{\bf F}_a\in\mathbb{Z}_p^{m\times d}$ time some vectors ${\bf x}_1,\dots,{\bf x}_b\in\mathbb{Z}_p^{d}$. While the cost of computing $a$ matrices multiplied by $b$ vectors is $\mathcal{O}(abmd)$ modular multiplications, using our scheme the client cost is $\mathcal{O}(am(b+d))$ modular multiplications and $\mathcal{O}(ad+bd+ab)$ modular exponentiations.

Fiore and Gennaro \cite{FG12} constructed  a multi-function verifiable
computation scheme for vector multiplication, it can be used to compute
matrix-vector multiplication by applying the solution to each row of the
matrix. When performing the same computations, our  scheme requires
less modular exponentiations compared with   \cite{FG12}.
 Experiments show that when the input matrices have $m$ rows, our scheme
is about $m$ times faster than the scheme in \cite{FG12}. Moreover,
 the running time of our scheme is less affected by the number of rows in
 the input matrix, while the cost of scheme in \cite{FG12} will increase
 linearly with the increase of the number of rows in the input matrices.
 Our scheme is more efficient both on the client side and on the server
 side.

\subsection
{Related Work}
In the cryptographic community, the idea  of  outsourcing expensive computations
has a long history.
The wallets with observers  of Chaum and Pedersen \cite{CP93} can be installed by a bank on the
client's  computer and assist the client to
do expensive computations. The wallets are not trusted by the client but still provide the assurance
  that they are  performing computations correctly by analyzing their communication with the bank.
  Hohenberger and Lysyanskaya \cite{HL05} presented protocols that allow  the client to offload
    the computation of modular exponentiations to two
non-colluding servers.
Golle and Mironov \cite{GM01} targeted on the the outsourcing of
inverting one-way functions.

The interactive proofs of \cite{Bab85,GMR85} allow a powerful prover to
show the truth of a statement to a weak verifier.
The probabilistically checkable proofs (PCPs)  of \cite{AS98}
allows the verifier to perform verification by checking only a few positions of
the entire proofs which however is too long for a weak verifier to process.
Kilian's efficient interactive arguments \cite{Kil92,Kil95} avoid the long proof with  a short commitment.
Micali's CS proofs \cite{Mic94} are non-interactive but require random oracles.

\vspace{1mm}
\noindent
{\bf Verifiable computation.}
The verifiable computation  of Gennaro et al. \cite{GGP10} gave a solution for the problem of  securely outsourcing computations, which  is both  non-interactive and  in the standard  model.
 The  verifiable computation schemes of
\cite{GGP10,CKV10,AIK10}
can delegate the functions as generic as any boolean circuits
 but have  very limited efficiency  due to the use of fully homomorphic encryption \cite{Gen09}.
 The {memory delegation}  \cite{CKLR11}
can  delegate  computations on an arbitrary
portion of the outsourced data. However, the
client  must be stateful and suffer from the efficiency issues  of
 PCP techniques. Benabbas et al.  \cite{BGV11} initiated a line of research on
 practical verifiable computation schemes for outsourcing specific
functions  such  as polynomials and matrices \cite{FG12,PST13}.
Parno et al. \cite{PRV12} initiated the study of public verifiable computation schemes.
 Both  \cite{PRV12} and \cite{FG12} proposed multi-function verifiable computation schemes
for different classes of functions.
The up to date implementations of efficient systems \cite{BCG13,BCG14,BFR13,CMT12,PHG13,SBV13,SMB12,SVP12,TJ13,TRM12,VSB13,WSR15} for verifiable
computations show that in this area  we are on the verge of achieving practical
efficiency.

\vspace{1mm}
\noindent
{\bf Homomorphic message authenticators.}
Homomorphic message authenticators \cite{GW13}  allow  one to
perform  certain admissible  computations over authenticated data and produce a
short tag that authenticates the result of the computation.
Using such schemes the client of a cloud service can securely outsource computations
on a set of authenticated data.
In the private-key setting, the homomorphic message authenticators, called homomorphic authentication
codes, have been constructed to admit linear functions \cite{AB09}, quadratic functions
\cite{BFR13}, and any polynomial functions
\cite{CF13}.
In the public-key setting,  the homomorphic message authenticators, called homomorphic signatures,
 have been  constructed to admit linear functions \cite{BFK09}, polynomial functions of bounded degrees
  \cite{BF11},  and any polynomial functions  \cite{GVW15}.
Some of them imply {outsourceable schemes \cite{BFR13}  while the others} only result in
schemes where the client-side computation is as heavy as the outsourced computation.

\vspace{1mm}
\noindent
{\bf Non-interactive proofs and arguments.}
 Goldwasser et al.  \cite{GKR08}  gave
 a  non-interactive scheme for delegating NC computations. However,
for any circuit of size $n$, the server's running time may be
a high degree polynomial of $n$ and  thus not practical.
The SNARGs or SNARKs of
\cite{BCCT12,GGPR13,BBFR15}
give non-interactive schemes for delegating  computations.
 However, they must rely on the
 non-falsifiable assumptions \cite{GW11}  which are both nonstandard   and
much stronger than the common assumptions  such as DDH.

\subsection {Application}

{\bf Digital Image Processing.}
In digital image processing \cite{RR06}, there are many ways to represent an image.
One of them is   using   2-D numerical arrays, which can be
described with matrices over a finite field, as long as the  field is large enough.
Each pixel of an image is a number and considered as an element
of its matrix representation.
An image with an $M\times N$ matrix representation can also be considered as
a column vector $\bf x$ with $d=MN$ entries.
Many useful operations in digital image processing such as
image restoration and image compression can be captured with
a linear transformation on the vector $\bf x$.
More precisely,
each of these operations can be realized by the multiplication
of an $m\times d$ matrix $\bf F$ with the column vector $\bf x$.
As the dimension of a digital image is typically very large,
the computation  of $\bf Fx$ is usually quite heavy.
When a weak client has multiple images ${\bf x}_1,{\bf x}_2,\ldots,{\bf x}_b$
and wishes to perform multiple operations ${\bf F}_1,{\bf F}_2,\ldots,
{\bf F}_a$ on these images,  our multi-matrix verifiable computation scheme would allow the client
to outsource the $ab$   matrix-vector multiplications to a powerful cloud server and
then  verify the server's results in a very fast way.

\vspace{1mm}
\noindent
{\bf Traffic Engineering.}
In traffic engineering \cite{CJ18}, traffic matrices may be used to describe  the traffic between the beginning
and the end of a  network. They are important tools   to plan
and manage the capacity of IP networks.
For example, one can derive a flow vector from the flow matrix which specifies
 the amount of traffic sent from a particular source to a particular
 destination, and obtain a link load vector by multiplying the routing
 matrix and the flow vector \cite{CJ18}. A large number of link load vectors
  are needed  in traffic engineering.
  That means a large number of multiplications between the routing matrices and
  the flow vectors must be performed.
  Our multi-matrix verifiable computation scheme allows the
  client to efficiently offload these computations to a cloud and
  also ensure the correctness of all computations with verification.

\vspace{1mm}
\noindent
{\bf Secure Distributed Computing.}
Our scheme  decouples
the process of preparing  functions and  the process of preparing   inputs in  outsourcing computations.
It allows the clients to distribute  many heavy computations (i.e., the matrix-vector multiplications) to multiple cloud servers and then  perform efficient verifications.
Compared with the model of   \cite{GGP10}, the main cost saving will stem from
the one-time preparation of each input, which is available for all functions.
Compared with   Fiore and Gennaro \cite{FG12}, the main cost saving
stems from the  batch verification of $m$ inner product computations, which
can significantly reduces the client's waiting time.

\subsection
{Organization}

The rest of the paper is organized as follow. In Section 2 we recall
the definition of multi-function verifiable computation. In Section 3 we present the
new multi-matrix verifiable computation scheme. In Section 4 we implement the
 new scheme and compare  with the multi-function scheme  of \cite{FG12}.
 Finally, Section 5 contains some concluding remarks.

\section{Model and Definition}

Multi-function  verifiable computation   \cite{PRV12,FG12,WL18} is a verifiable computation scheme
where the key generation process of encoding functions and the preparation of function inputs are decoupled such that
delegating the computation of multiple functions on multiple preprocessed function inputs is   possible.
Multi-function verifiable computation allows the client to  significantly reduce the
 time invested in the repeated work of preprocessing inputs
 such that the delegation becomes  outsourceable with  multiple functions.
Let $\cal F$ be a family of functions. Formally, a multi-function verifiable computation scheme
 $\Pi=({\sf Setup, KeyGen,ProbGen,Compute,}$\\ ${\sf Verify})$ for $\cal F$
consists of five probabilistic  polynomial-time algorithms, which can be defined as follows.
\begin{itemize}
\item
{${\sf Setup}(1^\lambda,{\cal F})\rightarrow(PK,SK)$:}
This is a {\em setup} algorithm that takes the  security parameter $\lambda$ and the function family  $\cal F$ as input. It
generates a set $PK$ of {\em public} parameters  and a set  $SK$ of {\em private} parameters. Both the public and the private parameters   will be used to
prepare  the functions and the inputs for delegation.

\item
{${\sf KeyGen}(PK,SK,f)\rightarrow(EK_f,VK_f)$:}
This is a {\em key generation} algorithm that takes the set $PK$ of public parameters, the set $SK$ of private parameters, and any function
$f\in {\cal F}$ as input. It produces both a public {\em evaluation key} $EK_f$, which will be used by the servers to
perform the delegated computations, and a {\em verification key} $VK_f$, which will be used by the client to verify
the server's work.

\item
{${\sf ProbGen}(PK,SK,x)\rightarrow(\sigma_x,VK_x)$:}
This is a {\em problem generation} algorithm that takes
the set $PK$ of public parameters, the set $SK$ of private parameters, and any function input
$x\in {\rm Dom}(f)$ as input.
It  produces both a  public {\em encoding} $\sigma_x$ of the input $x$, which will be used by the server to
perform the delegated computation,  and a {\em verification key} $VK_x$, which will be used by the client to
verify the server's work.
\item
{${\sf Compute}(EK_f,\sigma_x)\rightarrow \sigma_y$:}
This is the {\em server-side} algorithm that takes
the public evaluation key $EK_f$ and the public encoding of $x$ as input.
It computes and outputs an encoded version of the   value $y=f(x)$.

\item
${\sf Verify}(VK_f,VK_x,\sigma_y)\rightarrow \{f(x), \perp\}$:
This is a {\em verification} algorithm that takes the verification keys $VK_f, VK_x$ and the
server's computation result $\sigma_y$ as input. It determines whether $\sigma_y$
is a valid encoding of $f(x)$, and outputs  either $f(x)$ or $\perp$, where $\perp$
indicates that $\sigma_y$ is invalid.

\end{itemize}
A multi-function verifiable computation scheme is said to be {\em publicly delegatable} if the set
$SK$ of private parameters is empty such that any user of the scheme can run the
algorithms $\sf KeyGen$ and $\sf ProbGen$ to prepare its functions and/or inputs for delegation; otherwise, the
scheme is said to be {\em privately delegatable.} A multi-function verifiable computation scheme is said to be
{\em publicly verifiable} if the verification keys $VK_f$ and $VK_x$ can be made public such that
any entity can run the verification algorithm to verify if the server-side computation has been performed
correctly; otherwise, if  $VK_f$ and $VK_x$ must be kept secret, the scheme is said to be {\em privately
verifiable}. In this paper, we construct multi-verifiable computation schemes that are
publicly delegatable and privately verifiable.
The remaining definitions in the section will be given for the
privately verifiable setting.

A multi-function verifiable computation scheme is required  to be correct, secure and outsourceable.
Informally, a  multi-function verifiable computation  scheme     is said to be  {\em correct} if
the setup algorithm, the key generation algorithm and the problem generation algorithm
 produce  values that always enable
the honest servers to compute values that will
verify successfully and be converted into the correct function output $f(x)$.

\begin{definition}
\label{def:correct}
{\bf (correctness)}
Let $\cal F$ be a family of functions.
The multi-function verifiable computation  scheme $\Pi$ is  ${\cal F}$-{\em correct} if   for any
$(PK,SK)\leftarrow {\sf Setup}(1^\lambda,{\cal F})$,  any $f\in {\cal F}$,  any $x\in {\rm Dom}(f)$,
any  $(EK_f,VK_f)\leftarrow  {\sf KeyGen}(PK,SK,f)$, any $(\sigma_x,VK_x )\leftarrow {\sf ProbGen}(PK,SK,\\x)$,
and the faithfully computed server result
  $\sigma_y\leftarrow {\sf Compute}(EK_f,\sigma_x)$, it is always true that
${\sf Verify}(VK_f,\\ VK_x ,\sigma_y )=f(x)$.
\end{definition}

Informally, a multi-function  verifiable computation scheme is said to be     {\em secure} if no probabilistic polynomial-time
strategy of the  malicious
server can  persuade the verification
algorithm to accept a carefully crafted result, which
will cause the client to reconstruct   a   value $\hat{y}\neq f(x)$. This intuition can be formalized
by an experiment as bellow.\\
Experiment ${\bf Exp}_{\mathcal{A}}^{ {\rm PriVerif}}[\Pi,{\cal F},\lambda]$
\begin{itemize}
\item $(PK,SK)\leftarrow  {\sf Setup}(1^\lambda,{\cal F});$
\item $(f,x^*,\hat{\sigma_y})\leftarrow\mathcal{A}^{{\cal O}_{{\sf KeyGen}}(\cdot), ~{\cal O}_{\sf ProbGen}(\cdot),~
{\cal O}_{\sf Verify}(\cdot,\cdot,\cdot)}(PK);$
\item $\hat{y}\leftarrow {\sf Verify}(VK_f,VK_{x^*},\hat{\sigma_y});$
\item If $\hat{y}\neq\perp$ and $\hat{y}\neq f(x^*)$, output 1, else output 0.
\end{itemize}

In this experiment, a set $PK$ of public parameters and a set $SK$ of private parameters are firstly generated.
The adversary is   given access to three oracles
${\cal O}_{ {\sf KeyGen}}(\cdot),~ {\cal O}_{ {\sf ProbGen}}(\cdot)$ and ${\cal O}_{ {\sf Verify}}(\cdot )$, which
can be   defined as follows.
\begin{itemize}
\item ${\cal O}_{ {\sf KeyGen}}(\cdot)$:
On any input $f\in {\cal F}$, this oracle runs the key generation algorithm
 $ {\sf KeyGen}(PK,SK,f)$ to compute both a public evaluation key $EK_f$ and a verification key
  $VK_f$; it  returns
 $EK_f$ and stores $VK_f$.
 \item ${\cal O}_{ {\sf ProbGen}}(\cdot)$:
  On  input $x\in {\rm Dom}(f)$, this oracle
 runs the problem generation algorithm ${\sf ProbGen}(PK,SK,\\x)$ to compute both a
 public encoding $\sigma_x$ and a verification key $VK_x$;  it returns $\sigma_x$
 and stores $VK_x$.
 \item ${\cal O}_{ {\sf Verify}}(\cdot)$: On
  input $f\in {\cal F}$,  $x\in {\rm Dom}(f)$ and a purported output $\sigma_y$, this oracle
  runs   $ {\sf Verify}(VK_f,VK_x ,\\ \sigma_y )$ and returns the output of this algorithm.
\end{itemize}
After making a certain  number (polynomial in the security parameter $\lambda$) of queries to these oracles,
the adversary $\cal A$ carefully crafts a triple
$(f,x^*,\hat{\sigma}_y)$, where $f\in {\cal F}, x^*\in {\rm Dom}(f)$ and
$\hat{\sigma}_y$ is a purported output for the computation of $f(x^*)$, and
expects that the verification algorithm ${\sf Verify}(VK_f,VK_{x^*},\hat{\sigma}_y )$
will output a value   $\hat{y}\notin \{f(x^*),\perp\}$.
We say that the adversary $\cal A$ wins in the experiment
${\bf Exp}_{\mathcal{A}}^{ {\rm PriVerif}}[\Pi,{\cal F},\lambda]$ and define
${\bf Exp}_{\mathcal{A}}^{ {\rm PriVerif}}[\Pi,{\cal F},\lambda]=1$ if
$\hat{y}\notin \{f(x^*),\perp\}$.
For any security parameter $\lambda\in \mathbb{N}$, any function family $\cal F$, the {\em advantage}
of $\cal A$ making at most $q$ queries in the above experiment against $\Pi$ is defined as
\begin{equation*}
{\bf Adv}_{\cal A}^{\rm PriVerif}(\Pi, {\cal F}, q,\lambda)=\Pr[{\bf Exp}_{\mathcal{A}}^{ {\rm PriVerif}}[\Pi,{\cal F},\lambda]=1].
\end{equation*}
\begin{definition}
\label{def:correct}
{\bf (security)}
Let $\lambda$ be a security parameter and let
$\cal F$ be a family of functions.
The multi-function verifiable computation  scheme $\Pi$ is said to be
{\em $\cal F$-secure} if for any probabilistic polynomial-time adversary
$\cal A$, there is a negligible function $\rm negl(\cdot)$ such that
\begin{equation*}
{\bf Adv}_{\cal A}^{\rm PriVerif}(\Pi, {\cal F}, q,\lambda) \leq {\rm negl}(\lambda).
\end{equation*}
\end{definition}

In a multi-function verifiable computation,
we consider a scenario of computing $a$ different functions
$f_1,\ldots,\\ f_a\in {\cal F}$ on $b$ different function inputs
$x_1,\ldots,x_b$.
Informally, we say that a multi-function verifiable computation scheme is
{\em outsourceable} if  the total time cost
for encoding the functions, preparing the inputs and performing the verifications is
substantially less than the time cost of computing
all $ab$ results $\{f_i(x_j): i\in[a], j\in [b]\}$ from scratch.
\begin{definition}
\label{def:efficient}
{\bf (outsourceable)}
The multi-function verifiable computation scheme $\Pi$ is  {\em outsourceable} if it permits efficient
 generation, preparation,  verification and decoding.
That is, for any functions $f_1, \ldots, f_a\in {\cal F}$, any inputs
 $x_1,\dots,x_b$,  and any server results
$\sigma_{ij}$ for the computation of $f_i(x_j)$,  the total time required for
 $\{{\sf KeyGen}(PK,SK,f_i)\}_{i=1}^a$,
  $\{{\sf ProbGen}(PK,SK,x_j)\}_{j=1}^b$, $\{{\sf Verify}(VK_{f_i}, VK_{x_j}, \sigma_{ij}):i\in[a], j\in [b]\}$
   is  $o(T)$, where $T$ is the time required to compute all $ab$ function outputs
 $\{f_i(x_j): i\in[a], j\in [b]\}$ from scratch.
\end{definition}
We also work in the amortized model of \cite{GGP10,FG12}. This is reflected in the above definition as the delegation
 of multiple functions were considered.

\section{Multi-Matrix Delegation Scheme}

Let $\lambda$ be a security parameter. Let $p$ be a $\lambda$-bit prime and let
$\mathbb{Z}_p$ be the finite field of $p$ elements. Let $m,d>0$
be integers and ${\cal F}_{m,d}$ be the set of all $m\times d$ matrices over
the finite field $\mathbb{Z}_p$. For any matrix
$\textbf{F}\in {\cal F}_{m,d}$, we interpret $\bf F$ as a matrix function that
takes any (column) vector ${\bf x}\in \mathbb{Z}_p^d$ as input and outputs
a (column) vector ${\bf Fx}\in \mathbb{Z}_p^m$.
In this section we shall provide a multi-function   scheme
for delegating  the functions in ${\cal F}_{m,d}$. The proposed scheme will be
both
 publicly delegatable and privately verifiable.

When $m=1$, Fiore and Gennaro \cite{FG12} has proposed a multi-function scheme
${\cal VC}_{MultiF}$, which is   publicly delegatable and privately verifiable, for the
function family ${\cal F}_{1,d}$.
In  the scenario of computing $a$ functions from ${\cal F}_{1,d}$  on $b$
inputs from $\mathbb{Z}_p^d$, the scheme ${\cal VC}_{MultiF}$ would require the client to perform
as many as $ab$ verifications, where each verification is expensive and involves several exponentiations in a cyclic group of
prime order $p$. Our scheme is proposed to significantly reduce the client-side cost
in verification.

Let $\mathbb{G}=\langle g\rangle$ be a cyclic group of prime order $p$.
The public parameters of ${\cal VC}_{MultiF}$ consists of $d$ uniformly chosen
groups elements $R_1,\dots,R_d\in \mathbb{G}$. For any function
${\bf f}=(f_1,\ldots,f_d)\in {\cal F}_{1,d}$, the key generation is done by
computing a value $W_j=g^{\alpha f_j} R_j^k$ for every $j\in[d]$, where
$\alpha, k\in \mathbb{Z}_p$ are randomly chosen and serve as a private verification key.
For any input ${\bf x}=(x_1,\ldots,x_d)^\top \in \mathbb{Z}_p^d$,
the problem generation is done by  computing a verification key
$VK_{\bf x}=\prod_{j=1}^d R_j^{x_j}$.
The server-side algorithm computes both the function value
$y=\sum_{j=1}^d f_j x_j$ and a proof $V=\prod_{j=1}^d W_j^{x_j}$.
Finally, the verification is done by checking the equality
$V=g^{\alpha y} \cdot (VK_{\bf x})^k$. The security of the scheme follows from the following facts:
(1) given both $\bf f$ and $W_1,\ldots,W_d$, the uniformly chosen field element
$\alpha$ is kept pseudorandom; (2) a successful attack of the scheme requires
the server to carefully craft both a value
$\hat{y}\neq y$ and a proof $\hat{V}$ such that $\hat{V}=g^{\alpha \hat{y}} (VK_{\bf x})^k$;
(3)  the equality  essentially requires $V/\hat{V}=g^{\alpha(y-\hat{y})}$, which can
be satisfied only with a negligible probability.

In this section, we shall consider the delegation of functions of ${\cal F}_{m,d}$, with
emphasis on improving the efficiency of both the client-side computation and the
server-side computation. Let ${\bf F}\in {\cal F}_{m,d}$ be any matrix function and let
${\bf x}\in \mathbb{Z}_p^d$ be any input. While the delegation of
${\bf Fx}$ can be accomplished by considering the function $\bf F$ as $m$
functions from ${\cal F}_{1,d}$, one for each row of the matrix,
the client-side verification requires checking
$m$ different equalities, which may be costly for large $m$.
Our idea of speeding-up the verification is simple and done by combining
the $m$ rows of $\bf F$ as a single function in ${\cal F}_{1,d}$ and
perform the verification as in ${\cal VC}_{MultiF}$.
In particular, the combining work is done by choosing a vector
${\bf r}\leftarrow \mathbb{Z}_p^m$ uniformly and computing
the single function as ${\bf s}={\bf r}{\bf F}$.
The new scheme can be detailed as follows.

\begin{itemize}
\item
{${\sf Setup}(1^\lambda,{\cal F}_{m,d})$:}
This algorithm takes the security parameter $\lambda$ and the function family
${\cal F}_{m,d}$ as input.
It generates the description of a cyclic group $\mathbb{G}=\langle g\rangle$ of prime order $p$, where $g$
is a random generator of the group. It chooses
$d$ group elements  $R_1,\ldots,R_d\leftarrow \mathbb{G}$ uniformly at random.
The algorithm outputs a set  $SK=\perp$ of private parameters  and a set
$PK=(p,\mathbb{G},g,R_1,\ldots,R_d)$ of public parameters.
\item
{${\sf KeyGen}(PK,SK,{\bf F})$:} This algorithm takes
the set  $PK=(p,\mathbb{G},g,R_1,\ldots,R_d)$ of public parameters, the set $SK=\perp$ of private parameters,  and a function
 ${\bf F}\in {\cal F}_{m,d}$ as input.  It chooses
 $k\leftarrow\mathbb{Z}_p$,  ${\bf r}=(r_1,\ldots,r_m)\leftarrow \mathbb{Z}_p^m$,
 all uniformly
 and at random. It  computes ${\bf s}=(s_1,\dots,s_d)={\bf r}{\bf F}$, and computes $W_j=g^{s_j}\cdot R_j^k$
 for every $j\in[d]$. Let
 $W=(W_1,\dots,W_d)$. This algorithm finally outputs a public evaluation key
  $EK_{\bf F}=({\bf F},W)$ and a private verification key $VK_{\bf F}=(k,{\bf r})$.
\item
{${\sf ProbGen}(PK,SK,{\bf x})$:}
This algorithm takes
the set  $PK=(p,\mathbb{G},g,R_1,\dots,R_d)$ of public parameters, the set $SK=\perp$ of private parameters,  and
any function input ${\bf x}=(x_1,\dots,x_d)^\top\in \mathbb{Z}_p^d$ as input.
It computes $VK_{\bf x}=\prod_{j=1}^dR_j^{x_j} $, outputs a public encoding
$\sigma_{\bf x}={\bf x}$ and  the public verification  key $VK_{\bf x}$.
\item
{${\sf Compute}(EK_{\bf F},\sigma_{\bf x})$:}
This algorithm takes the public evaluation key
$EK_{\bf F}=({\bf F},W)$ and the public encoding $\sigma_{\bf x}={\bf x}=(x_1,\ldots,x_d)^\top$
as input.
It computes $ {\bf y}=(y_1,\dots,y_m)^\top={\bf Fx} $ and $V=\prod_{j=1}^dW_j^{x_j}$. This algortihm
 outputs $\sigma_{\bf y}=({\bf y} ,V)$.
\item
{${\sf Verify}(VK_{\bf F},VK_{\bf x},\sigma_{\bf y})$:}
This algorithm takes the private verification key $VK_{\bf F}=(k, {\bf r})$, the verification key
$VK_{\bf x}=\prod_{j=1}^dR_j^{x_j} $ and the server's results $\sigma_{\bf y}=({\bf y},V)$
as input. If  $V=g^{{\bf ry}}\cdot (VK_{\bf x})^k$, this algorithm outputs ${\bf y}$;
 otherwise, it outputs $\perp$.
\end{itemize}

\subsection{Correctness}
The correctness of the scheme requires that for any $(PK,SK)\leftarrow {\sf Setup}(1^\lambda,{\cal F}_{m,d})$,
any function ${\bf F}\in {\cal F}_{m,d}$, any function input ${\bf x}\in \mathbb{Z}_p^d$, for any
$(EK_{\bf F},VK_{\bf F})\leftarrow {\sf KeyGen}(PK,SK,{\bf F})$, for any $(\sigma_{\bf x}, VK_{\bf x})\leftarrow
{\sf ProbGen}(PK,\\SK, {\bf x})$, if $\sigma_{\bf y}$ is faithfully computed by
executing the algorithm ${\sf Compute}(EK_{\bf F},\sigma_{\bf x})$, then it must be true that
${\sf Verify}(VK_{\bf F},VK_{\bf x},\sigma_{\bf y})={\bf Fx}$.
For our construction, it suffices to show that the equation $V=g^{  {\bf ry}} (VK_{\bf x})^k$
will be satisfied, as that will cause the client to output ${\bf y}={\bf Fx}$.
The equality can be proved as follows:
\begin{equation*}
\begin{split}	
V  &= \prod_{j=1}^dW_j^{x_j} \\ &= \prod_{j=1}^d\left(g^{s_j}\cdot R_j^k\right)^{x_j}\\
&= g^{  \sum_{j=1}^d s_j x_j}\cdot\left(\prod_{j=1}^dR_j^{x_j}\right)^k \\
 &= g^{ {\bf ry}}\cdot (VK_{\bf x})^k.
\end{split}
\end{equation*}
\subsection{Security}
 The security of the scheme $\Pi$ requires that no probabilistic polynomial-time adversary
 should be able to  persuade the client to accept a carefully crafted server result
 $\hat{\sigma}_{\bf y}$, which  will cause the client to output a wrong
 function value $\hat{\bf y}\neq {\bf Fx}$. Formally, this requires that any PPT adversary
 will succeed in the standard security experiment  ${\bf Exp}_{\mathcal{A}}^{ {\rm PriVerif}}[\Pi,{\cal F},\lambda]$
with at most a negligible advantage.
 In \cite{FG12} it was shown that the two-input function
 $H: \mathbb{Z}_p\times \mathbb{G}\rightarrow \mathbb{G}$ defined by
 $H_k(X)=X^k$ is a
 weak pseudorandom function such that  for any  PPT adversary
 $\cal A$ and any polynomial function  $d=d(\lambda)$, the advantage
\begin{equation*}
\begin{split}
\epsilon_{\rm wprf}:&=\left|\Pr[{\cal A}(\{(X_j,Y_j)\}_{j=1}^d)=1]\right.\\&-\left.\Pr[{\cal A}(\{(X_j,Z_j)\}_{j=1}^d)=1]\right|
\end{split}
\end{equation*}
of $\cal A$ distinguishing between the output distribution  of
$H$ on a set of randomly chosen group elements  and the uniform distribution is negligible in
$\lambda$,     where
 the probabilities are taken over
 $k\leftarrow \mathbb{Z}_p, \{X_j\}_{i=1}^d\leftarrow \mathbb{G}^d,
 \{Y_j\}_{j=1}^d=\{H_k(X_j)\}_{j=1}^d$ and $\{Z_j\}_{j=1}^d\leftarrow \mathbb{G}^d$.
 In our multi-matrix verifiable computation scheme  the weak PRF $H$ was also used in the computation of
 $W_j$ as $W_j=g^{s_j}\cdot H_k(R_j)$ for every $j\in[d]$.
\begin{theorem}
Any adversary $\mathcal{A}$ making  at most $q$
queries to the oracle $\mathcal{O}_{{\sf Verify}(\cdot,\cdot,\cdot)}$
 in the  experiment ${\bf Exp}_{\mathcal{A}}^{ {\rm PriVerif}}[\Pi,\\{\cal F},\lambda]$
 will succeed with probability  at most
$q\cdot \epsilon_{\rm wprf}+\frac{q}{p-q+1}$, i.e.,
 ${\bf Adv}_{\mathcal{A}}^{ {\rm PriVerif}}[\Pi,{\cal F},q,\lambda]\leq q\cdot \epsilon_{\rm wprf}+ \frac{q}{p-q+1}$.
In particular, if $q$ is a polynomial function of $\lambda$ and $p$ is a $\lambda$-bit prime, then
the adversary $\cal A$ succeeds with negligible probability.
\end{theorem}
\begin{proof}
In order to show that ${\bf Adv}_{\mathcal{A}}^{ {\rm PriVerif}}
[\Pi,{\cal F},q,\lambda]\leq q\cdot \epsilon_{\rm wprf} +\frac{q}{p-q+1}$, we define the
 following security experiments $E_0, E_1,  E_{2,0},\ldots, E_{2,q}, E_3$
 and denote by $E_0({\cal A}),\\ E_1({\cal A}), E_{2,0}({\cal A}),\dots,E_{2,q}({\cal A}),E_3({\cal A})$
 the events that $\cal A$ succeeds in the respective experiments, i.e., the events that the respective experiments output 1.

\vspace{2mm}
 \noindent
{\bf Experiment $E_0$}: This is the standard security experiment
  ${\bf Exp}_{\mathcal{A}}^{ {\rm PriVerif}}[\Pi,{\cal F},\lambda]$.

\vspace{2mm}
 \noindent
{\bf Experiment $E_1$}:
 This experiment is identical to $E_0$ except the  following changes.
Whenever the adversary $\cal A$ makes a query  $({\bf F},{\bf x}, \sigma_{\bf y})$  to the oracle
${\cal O}_{\sf Verify}(\cdot,\cdot,\cdot)$, where $VK_{\bf F}=(k,{\bf r}), VK_{\bf x}=\prod_{j=1}^d R_j^{x_j},
\sigma_{\bf y}=({\bf y},V)$,
the challenger performs the verification by checking  the equality
$V=g^{{\bf ry}}\cdot \prod_{j=1}^d H_k(R_j)^{x_j}$, instead of checking
the equality
$V=g^{{\bf ry}}\cdot (\prod_{j=1}^d R_j^{x_j})^k.$

\vspace{2mm}
 \noindent
{\bf Experiment $E_{2,i}$}:  For every integer $i=0,1,\ldots,q$, the experiment
 $E_{2,i}$  is identical to $E_1$  except the following changes to the first $i$ queries made by the adversary:
\begin{itemize}
\item
whenever $\cal A$ makes a query $\bf F$ to the oracle ${\cal O}_{\sf KeyGen}(\cdot)$, instead of choosing
$k\leftarrow \mathbb{Z}_p$  and computing each $W_j$ as $W_j=g^{s_j}\cdot R_j^k$, the challenger
chooses $d$ group elements $Z_1,\ldots, Z_d\leftarrow\mathbb{G}$,  computes
each $W_j$ as $W_j=g^{s_j}\cdot Z_j$, and keeps $VK_{\bf F}=(Z_1,\ldots,Z_d, {\bf r})$
for the purpose of  verification;

\item
whenever  $\cal A$ makes a query  $({\bf F},{\bf x}, \sigma_{\bf y})$  to the oracle
${\cal O}_{\sf Verify}(\cdot,\cdot,\cdot)$, where $\sigma_{\bf y}=({\bf y},V)$,
the challenger retrieves
$VK_{\bf F}=(Z_1,\ldots, Z_d, {\bf r})$ and  performs the verification by checking  the equality
$V=g^{{\bf ry}}\cdot \prod_{j=1}^d Z_j^{x_j}$.
\end{itemize}
It is  straightforward to see that the experiment $E_{2,0}$ is identical to $E_1$.

\vspace{2mm}
 \noindent
{\bf Experiment $E_3$}: This experiment is  the renaming of the experiment $E_{2,q}$.

It is easy to see that the change of $E_1$ with respect to $E_0$ has no impact on the probability that
$\cal A$ successfully breaks the security of the underlying scheme, i.e.,
\begin{equation}
\label{eqn:sec1}
\Pr[E_0({\cal A})]=\Pr[E_1({\cal A}) ]=\Pr[E_{2,0}({\cal A}) ].
\end{equation}
For every $i\in [q]$, the experiment $E_{2,i}$ is identical to $E_{2,i-1}$ except that
in the $i$th query the values of a  weak PRF $H_k$ in $\sf KeyGen$ and $\sf Verify$
 is replaced with the truly random group elements.
 We must have that
 \begin{equation}
 \label{eqn:sec2}
| \Pr[E_{2,i-1}({\cal A})]- \Pr[E_{2,i}({\cal A})]|\leq \epsilon_{\rm wprf}
 \end{equation}
 for every $i\in [q]$, because otherwise one would be able to distinguish between
the weak PRF and a truly random function with advantage  $>\epsilon_{\rm wprf}$, which however
 gives a contradiction.
 It remains to show that
 \begin{equation}
  \label{eqn:sec3}
\Pr[E_3({\cal A})=1]\leq \frac{q}{p-q+1},
 \end{equation}
 which together with (\ref{eqn:sec1}) and (\ref{eqn:sec2}) will give the expected conclusion, i.e.,
\begin{equation*}
{\bf Adv}_{\mathcal{A}}^{ {\rm PriVerif}}[\Pi,{\cal F},q,\lambda]\leq q\cdot \epsilon_{\rm wprf}+ \frac{q}{p-q+1}.
\end{equation*}

In the experiment $E_3$, the adversary $\cal A$  makes at most
$q$ queries to  the oracles.
Suppose that $\cal A$ has made a query to ${\cal O}_{\sf KeyGen}(\cdot)$ with $\bf F$.
Then the challenger would have chosen  $Z_1,\ldots,Z_d\leftarrow \mathbb{G}$,
chosen ${\bf r}\leftarrow \mathbb{Z}_p^m$, computed  ${\bf s}=(s_1,\ldots,s_d)={\bf r}{\bf F}$,
computed  $W_j=g^{s_j} Z_j$ for every $j\in[d]$, and kept $VK_{\bf F}=(Z_1,\ldots,Z_d,{\bf r})$
for verification.
Whenever $\cal A$ makes a query to ${\cal O}_{\sf Verify}(\cdot,\cdot,\cdot)$ with $({\bf F},{\bf x}, \hat{\sigma})$,
 where $\hat{\sigma}=(\hat{\bf y}, \hat{V})$, the challenger would verify
 if
\begin{equation}\label{eqn:forge}
\hat{V}=g^{{\bf r}\hat{\bf y}}\prod_{j=1}^d Z_j^{x_j}.
\end{equation}
The  query $({\bf F},{\bf x}, \hat{\sigma})$ allows $\cal A$ to win in $E_3$ if and only if
$\hat{\bf y}\neq {\bf Fx}$ but the equality
(\ref{eqn:forge}) still holds.
On the other hand,
let ${\bf y}={\bf Fx}$ and  $V=\prod_{j=1}^d W_j^{x_j}$ be the response that would be computed by
an honest server. The correctness of the scheme would imply that
\begin{equation*}
 {V}=g^{{\bf r} {\bf y}}\prod_{j=1}^d Z_j^{x_j}.
\end{equation*}
As a result, the query $({\bf F},{\bf x}, \hat{\sigma})$ allows $\cal A$ to win in $E_3$
if and only if
\begin{equation}
\label{eqn:forge1}
(\hat{\bf y}\neq {\bf y})\wedge (\hat{V}/V=g^{{\bf r}(\hat{\bf y}-{\bf y})}).
\end{equation}
For every $\ell\in [q]$, we denote by $S_\ell$ the event that (\ref{eqn:forge1}) is satisfied
 in the $\ell$-th query to ${\cal O}_{\sf Verify}(\cdot,\cdot,\cdot)$.
 Then it is easy to see that $E_3({\cal A})$ occurs if and only if for at least one  of the
 $\ell\in [q]$, the event $S_\ell$ occurs.
 Then we would have that
  \begin{equation}
 \label{eqn:in1}
 \begin{split}
 \Pr[E_3({\cal A})]&=\Pr[\vee_{\ell=1}^q S_\ell]\\
 &\leq \Pr[S_1]+\sum_{\ell=2}^q\Pr[S_\ell|\wedge_{i=1}^{\ell-1}\bar{S}_i ],
 \end{split}
 \end{equation}
 where the inequality is  a standard result from discrete probability theory.

It is not hard to see that the adversary $\cal A$ learns absolutely no information about
  $\bf r$ from the queries to
 ${\cal O}_{\sf KeyGen}(\cdot)$ in  $E_3$.  In fact,  the oracle's answer $({\bf F},W)$
is completely independent of $\bf r$ because each $W_j$ was computed as
$W_j=g^{s_j}\cdot Z_j$ and the $Z_j$ was chosen uniformly at random and independent everything else in the experiment.
On the other hand, it is also easy to see that the adversary $\cal A$ learns absolutely no information about
$\bf r$ from the queries to  ${\cal O}_{\sf ProbGen}(\cdot)$.
This is because the oracle's answer $VK_{\bf x}$ for each $\bf x$ was computed as
$VK_{\bf x}=\prod_{j=1}^d R_j^{x_j}$, which is completely independent of
$\bf r$. Therefore, before making any queries to ${\cal O}_{\sf Verify}(\cdot,\cdot,\cdot)$
the verification key $\bf r$ for each function is still uniformly distributed over
$\mathbb{Z}_p^m$, from the point of view of $\cal A$.

Each query   $({\bf F},{\bf x}, (\hat{\bf y}, \hat{V}))$ to ${\cal O}_{\sf Verify}(\cdot,\cdot,\cdot)$
with $\hat{\bf y}\neq {\bf y}$ would either allow the adversary $\cal A$ to win
in $E_3$ (when $\hat{V}/V=g^{{\bf r}(\hat{\bf y}-{\bf y})}$) or give some information about
$\bf r$ to $\cal A$ (when $\hat{V}/V\neq g^{{\bf r}(\hat{\bf y}-{\bf y})}$).
The former event will occur if and only if $\bf r$ happens to a solution of the following equation system
\begin{equation}
\label{eqn:sys}
(\hat{\bf y}-{\bf y}) {\bf r}=\log_g (\hat{V}/V),
\end{equation}
where $\log_g (\hat{V}/V)$ is the discrete logarithm of
$\hat{V}/V\in \mathbb{G}$ with respect to the group generator $g\in \mathbb{G}$.
The latter event will give $\cal A$  at most  the knowledge that ${\bf r}$ is not
a solution of the equation system (\ref{eqn:sys}), which can be realized only if
$\cal A$ has chosen $\hat{V}$ in a special way (for example, by choosing $\hat{v}\in \mathbb{Z}_p$ and setting
$\hat{V}=V\cdot g^{\hat{v}}$).

When the first query was being made to ${\cal O}_{\sf Verify}(\cdot,\cdot,\cdot)$,
the $\bf r$ was uniformly distributed over the set $\mathbb{Z}_p^m$.
No matter which $\hat{\bf y}\neq {\bf y}$ was chosen by $\cal A$,
the equation system (\ref{eqn:sys}) will have $p^{m-1}$ solutions in $\mathbb{Z}_p^m$.
As a result,      the uniformly distributed  $\bf r$
will happen to be a solution of
(\ref{eqn:sys}) with probability $\epsilon_1=p^{m-1}/p^m=1/p$.
In general, for every $\ell\in [q]$, if $\ell-1$ queries
have been made such that either $\hat{\bf y}={\bf y}$ or
(\ref{eqn:sys}) was not satisfied, then
each such query would allow $\cal A$ to rule out at most
$p^{m-1}$ possibilities of $\bf r$ over the set $\mathbb{Z}_p^m$.
Therefore, conditioned on $\bar{S}_1\wedge\cdots\wedge \bar{S}_{\ell-1}$,
the private key {\bf r} should be uniformly distributed over a  subset  of $\mathbb{Z}_p^m$
of $\geq p^m-(\ell-1)p^{m-1}$ elements. It follows that
\begin{equation}
\label{eqn:in2}
\begin{split}
\Pr[S_\ell|\wedge_{i=1}^{\ell-1}\bar{S}_i]
&\leq\frac{p^{m-1}}{p^m-(\ell-1)p^{m-1}}
\\&
=\frac{1}{p-(\ell-1)}
\end{split}
\end{equation}
for every $\ell\in [q]$. The equalities (\ref{eqn:in1}) and (\ref{eqn:in2})
imply that
{
\begin{equation*}
\begin{split}
\Pr[E_3({\cal A})]&\leq \sum_{\ell=1}^q \Pr[S_\ell|\wedge_{i=1}^{\ell-1}\bar{S}_i] \\
& \leq \sum_{\ell=1}^q \frac{1}{p-(\ell-1)}\\
&\leq\frac{q}{p-q+1},
\end{split}
\end{equation*}
}
which  gives the expected inequality (\ref{eqn:sec3}).
\qed
\end{proof}

{\color{black}

\section{Performance Analysis}
In this section, we consider the scenario  of
outsourcing the multiplications
of $a$ matrices ${\bf F}_1,{\bf F}_2,\dots,{\bf F}_a \in {\cal F}_{m,d}$
with $b$ vectors ${\bf x}_1,{\bf x}_2,\dots,{\bf x}_b \in \mathbb{Z}_p^m$.
We shall evaluate our multi-matrix verifiable computation scheme with
several  complexity measures, such as the computation complexity,
the communication complexity and the storage complexity.
The evaluations will be done both in theory and with experiments.
We   show that the multi-function scheme of \cite{FG12}
is a special case of ours for $m=1$ and our scheme
will be substantially  more efficient than \cite{FG12} for large $m$.

\subsection{Theoretical Analysis}
\label{sec:ta}

{\bf Computation Complexity.}
In our scheme, the algorithm    ${\sf Setup}(1^\lambda, {\cal F}_{m,d})$
chooses $d$ random elements from $\mathbb{G}$, a cyclic group  of $p$ elements.
For every ${\bf F}\in \{{\bf F}_1,{\bf F}_2, \\
\ldots, {\bf F}_a\}$, the execution of
${\sf KeyGen}(PK,SK,{\bf F})$
requires the client to choose $k\leftarrow \mathbb{Z}_p,{\bf r}\leftarrow
\mathbb{Z}_p^m$, compute ${\bf s}={\bf rF}$, and
$W_j=g^{s_j}R_j^k$ for every $j\in[d]$.
Each execution consists of  $m+1$ random number generations,
$(m-1)d$ additions modulo $p$, $md$ multiplications modulo $p$,
$2d$ exponentiations in $\mathbb{G}$, and $d$ multiplications in $\mathbb{G}$.
For every ${\bf x}\in \{{\bf x}_1,{\bf x}_2,\ldots,{\bf x}_b\}$,
the execution of ${\sf ProbGen}(PK,SK,{\bf x})$
requires the client to compute
     $VK_{\bf x}=\prod_{i=1}^dR_i^{x_i}$.
     The execution consists of
     $d$ exponentiations in $\mathbb{G}$
     and $d-1$ multiplications in
      $\mathbb{G}$.
     For every ${\bf F}\in \{{\bf F}_1,{\bf F}_2,
\ldots, {\bf F}_a\}$ and
${\bf x}\in \{{\bf x}_1,{\bf x}_2,\ldots,{\bf x}_b\}$,
${\sf Compute}(EK_{\bf F},\sigma_{\bf x})$ requires the server
to compute both the result  ${\bf y}={\bf Fx}$
and a proof  $V=\prod_{i=1}^dW_i^{x_i}$.
The execution consists of $m(d-1)$ additions modulo $p$,
$md$ multiplications modulo $p$, $d$ exponentiations
in $\mathbb{G}$ and $d-1$ multiplications in $\mathbb{G}$.
 For every ${\bf F}\in \{{\bf F}_1,{\bf F}_2,
\ldots, {\bf F}_a\}$ and
${\bf x}\in \{{\bf x}_1,{\bf x}_2,\ldots,{\bf x}_b\}$,
 ${\sf Verify}(VK_{\bf F},VK_{\bf x},\sigma_{\bf y})$
requires the client to  verify if   $V=g^{{\bf ry}}\cdot (VK_{\bf x})^k$.
The execution consists of $m-1$ additions modulo $p$,
$m$ multiplications modulo $p$, 1 multiplication  in $\mathbb{G}$
and 2 exponentiations in $\mathbb{G}$.

\begin{threeparttable}[!htbp]
\center\caption{Computation Complexity}\label{Table1}
\small
\begin{tabular}{|c|ccccc|}
\hline
{Algorithm}&{\bf rng}&${\bf add}_p$&${\bf mul}_p$&${\bf mul}_\mathbb{G}$&${\bf exp}_\mathbb{G}$\\
\hline
\multirow{2}{*}{\sf Setup}&$d$&0&0&0&0\\
&\cellcolor[HTML]{C0C0C0}$d$&\cellcolor[HTML]{C0C0C0}0&\cellcolor[HTML]{C0C0C0}0&\cellcolor[HTML]{C0C0C0}0&\cellcolor[HTML]{C0C0C0}0\\
\hline
\multirow{2}{*}{\sf KeyGen}&\!\!$a(m\!+\!1)$\!\!&\!\!$a(m\!-\!1)d$\!\!&\!$amd$\!&$ad$&$2ad$\\
&\cellcolor[HTML]{C0C0C0}$2am$&\cellcolor[HTML]{C0C0C0}0&\cellcolor[HTML]{C0C0C0}\!\!$amd$\!\!&\cellcolor[HTML]{C0C0C0}$amd$&\cellcolor[HTML]{C0C0C0}\!\!$2amd$\!\!\!\\
\hline
\multirow{2}{*}{\sf ProbGen}&0&0&0&\!$b(d\!-\!1)$\!&$bd$\\
&\cellcolor[HTML]{C0C0C0}0&\cellcolor[HTML]{C0C0C0}0&\cellcolor[HTML]{C0C0C0}0&\cellcolor[HTML]{C0C0C0}$\!b(d\!-\!1)$\!&\cellcolor[HTML]{C0C0C0}$bd$\\
\hline
\multirow{2}{*}{\sf Compute}&0&\!\!$abm(d\!-\!1)$\!\!&\!\!$abmd$\!\!&\!\!$ab(d\!-\!1)$\!\!&$abd$\\
&\cellcolor[HTML]{C0C0C0}0&\cellcolor[HTML]{C0C0C0}\!\!$abm(d\!-\!1)$\!\!&\cellcolor[HTML]{C0C0C0}\!\!$abmd$\!\!&\cellcolor[HTML]{C0C0C0}\!\!$abm(d\!-\!1)$\!\!&\cellcolor[HTML]{C0C0C0}\!\!$abmd$\!\!\!\\
\hline
\multirow{2}{*}{\sf Verify}&0&\!\!$ab(m\!-\!1)$\!\!&\!$abm$\!&$ab$&$2ab$\\
&\cellcolor[HTML]{C0C0C0}0&\cellcolor[HTML]{C0C0C0}0&\cellcolor[HTML]{C0C0C0}\!$abm$\!&\cellcolor[HTML]{C0C0C0}$abm$&\cellcolor[HTML]{C0C0C0}\!\!$2abm$\!\!\!\\
\hline
\end{tabular}
\begin{itemize}
  \item  non-shaded numbers: our computation complexity
\item shaded numbers: computation complexity of \cite{FG12}
  \item{\bf rng}:  random number generation
  \item${\bf add}_p$:   addition modulo $p$
  \item${\bf mul}_p$:   multiplication  modulo $p$
  \item${\bf mul}_\mathbb{G}$:   multiplication in $\mathbb{G}$
  \item${\bf exp}_\mathbb{G}$:  exponentiation in $\mathbb{G}$
\end{itemize}
\end{threeparttable}

\vspace{2mm}
\noindent
Table \ref{Table1} provides both a summary of the above analysis and  comparisons between our scheme
and   \cite{FG12} for  outsourcing  the $ab$ computations $\{{\bf F}_i {\bf x}_j: i\in[a],j\in[b]\}$.
In particular, the non-shaded numbers  describe  our scheme and
the  shaded numbers describe  \cite{FG12}.
As   \cite{FG12}
is designed for computing the inner product  of   two vectors,
in Table 1
the shaded numbers    are obtained by executing
the scheme of \cite{FG12}   for  $abm$ inner product computations.
{\color{black}
We  denote with $t_{\bf rng}, t_{{\bf add}_p},
t_{{\bf mul}_p},  t_{{\bf mul}_{\mathbb{G}}}, $
and $ t_{{\bf exp}_{\mathbb{G}}}$ the time required by each of the operations
 ${\bf rng},  {{\bf add}_p},
 {{\bf mul}_p},  {{\bf mul}_{\mathbb{G}}}, $
and $  {{\bf exp}_{\mathbb{G}}}$, respectively.
We denote with
$t_{\rm c}^1$ (resp. $t_{\rm c}^2$) and $t_{\rm s}^1$ (resp. $t_{\rm s}^2$)
the client-side computation time and the server-side computation time in
our scheme (resp. the scheme of [26]).  Then   Table \ref{Table1}  shows that
\begin{equation*}
\begin{split}
t_{\rm c}^1=&a(m+1)\cdot t_{\bf rng}+
a(m-1)(b+d)\cdot t_{{\bf add}_p}+\\
&am(b+d)\cdot t_{{\bf mul}_p}+
(ad+b(d-1)+ab)\cdot t_{{\bf mul}_{\mathbb{G}}}+\\
&(2ad+bd+2ab)\cdot t_{{\bf exp}_{\mathbb{G}}};  \\
t_{\rm c}^2=&2am\cdot t_{{\bf rng}}+am(b+d)\cdot t_{{\bf mul}_p}+ \\
&(amd+b(d-1)+abm)\cdot t_{{\bf mul}_{\mathbb{G}}}+\\
&(2amd+bd+2abm)\cdot t_{{\bf exp}_{\mathbb{G}}};\\
t_{\rm s}^1=&abm(d-1)\cdot t_{{\bf add}_p}+
abmd\cdot t_{{\bf mul}_p}+\\
&ab(d-1)\cdot t_{{\bf mul}_{\mathbb{G}}}+
abd\cdot t_{{\bf exp}_{\mathbb{G}}};   \\
t_{\rm s}^2=&abm(d-1)\cdot t_{{\bf add}_p}+abmd\cdot t_{{\bf mul}_p}+ \\
&abm(d-1)\cdot t_{{\bf mul}_{\mathbb{G}}}+abmd\cdot t_{{\bf exp}_{\mathbb{G}}}.
\end{split}
\end{equation*}

\vspace{2mm}
\noindent
It's easy to see that we always have
$t_{\rm c}^2\geq t_{\rm c}^1$ and $t_{\rm s}^2\geq t_{\rm s}^1$, i.e.,
our scheme is always faster than \cite{FG12},
 in terms of both client-side computation and server-side computation.
In particular, when $a=b=m$,  $d\rightarrow \infty$, and
$t_{{\bf exp}_{\mathbb{G}}}\gg
\max\{t_{{\bf mul}_{\mathbb{G}}},mt_{{\bf mul}_p},
mt_{{\bf add}_p}\}$, we will have
\begin{equation}
\label{eqn:time}
\begin{split}
t_{\rm c}^2/t_{\rm c}^1\geq
2m/3;   \hspace{5mm}
t_{\rm s}^2/t_{\rm s}^1\approx m.
\end{split}
\end{equation}
}

\vspace{2mm}
\noindent
{\bf Communication Complexity.}
 For every function ${\bf F}\in \{{\bf F}_1,{\bf F}_2,
\ldots, {\bf F}_a\}$ and every input
${\bf x}\in \{{\bf x}_1,{\bf x}_2,\ldots,{\bf x}_b\}$,
 our scheme   requires the client
to  send $EK_{\bf F},\sigma_{\bf x}$  to the server
and receive
$\sigma_{\bf y}$ from  the server. In our scheme,
 $EK_{\bf F}=({\bf F},W)$ consists of $md$ elements in $\mathbb{Z}_p$
 and $d$ elements in $\mathbb{G}$, $\sigma_{\bf x}$  consists of $d$
 elements in $\mathbb{Z}_p$, and $\sigma_{\bf y}=({\bf y},V)$ consists of
  $m$ elements in $\mathbb{Z}_p$ and one element in $\mathbb{G}$.

 \vspace{-2mm}

\begin{threeparttable}[!htbp]
\center\caption{Communication Complexity}\label{Table2}
\small
\setlength{\tabcolsep}{5mm}{
\begin{tabular}{|c|c|c|}
\hline
&Elements in $\mathbb{Z}_p$&Elements in $\mathbb{G}$\\
\hline
\multirow{2}{*}{$EK_{\bf F}$}&$amd$&$ad$\\
&\cellcolor[HTML]{C0C0C0}$amd$&\cellcolor[HTML]{C0C0C0}$amd$\\
\hline
\multirow{2}{*}{$\sigma_{\bf x}$}&$bd$&0\\
&\cellcolor[HTML]{C0C0C0}$bd$&\cellcolor[HTML]{C0C0C0}0\\
\hline
\multirow{2}{*}{$\sigma_{\bf y}$}&$abm$&$ab$\\
&\cellcolor[HTML]{C0C0C0}$abm$&\cellcolor[HTML]{C0C0C0}$abm$\\
\hline
\end{tabular}}
\begin{itemize}
  \item   non-shaded numbers: our communication complexity
    \item shaded numbers: communication  complexity of \cite{FG12}
\end{itemize}
\end{threeparttable}

\vspace{2mm}
\noindent
Table \ref{Table2} provides both a summary of the above analysis and  comparisons between our scheme
and   \cite{FG12} for  outsourcing the $ab$ computations $\{{\bf F}_i {\bf x}_j: i\in[a],j\in[b]\}$.
In particular, the non-shaded numbers   describe  our scheme   and
the   shaded numbers  describe  \cite{FG12}.
{\color{black}
We  denote with $\ell_p$ (resp.  $\ell_{\mathbb{G}}$)
the length in bits of each element of $\mathbb{Z}_p$ (resp. $\mathbb{G}$).
We denote with
$c^1$ (resp. $c^2$) the communication complexity of
our scheme (resp. \cite{FG12}). Then Table \ref{Table2} shows that
\begin{equation*}
\begin{split}
c^1&=(amd+bd+abm)\ell_p+(ad+ab)\ell_{\mathbb{G}};\\
c^2&=(amd+bd+abm)\ell_p+(amd+amb)\ell_{\mathbb{G}}.
\end{split}
\end{equation*}
It's easy to see that $c^1<c^2$, i.e., the communication complexity of our scheme is
always  lower than
  \cite{FG12}.
In particular,  when $\ell_p=O(\ell_{\mathbb{G}})$ and $amd+amb\gg ad+ab+bd$, we
will have
\begin{equation}
\label{eqn:comm}
c^2/c^1\approx 1+ \ell_{\mathbb{G}}/\ell_p.
\end{equation}

}

\begin{figure*}[h]
\subfigure{
\includegraphics[width=0.23\textwidth]{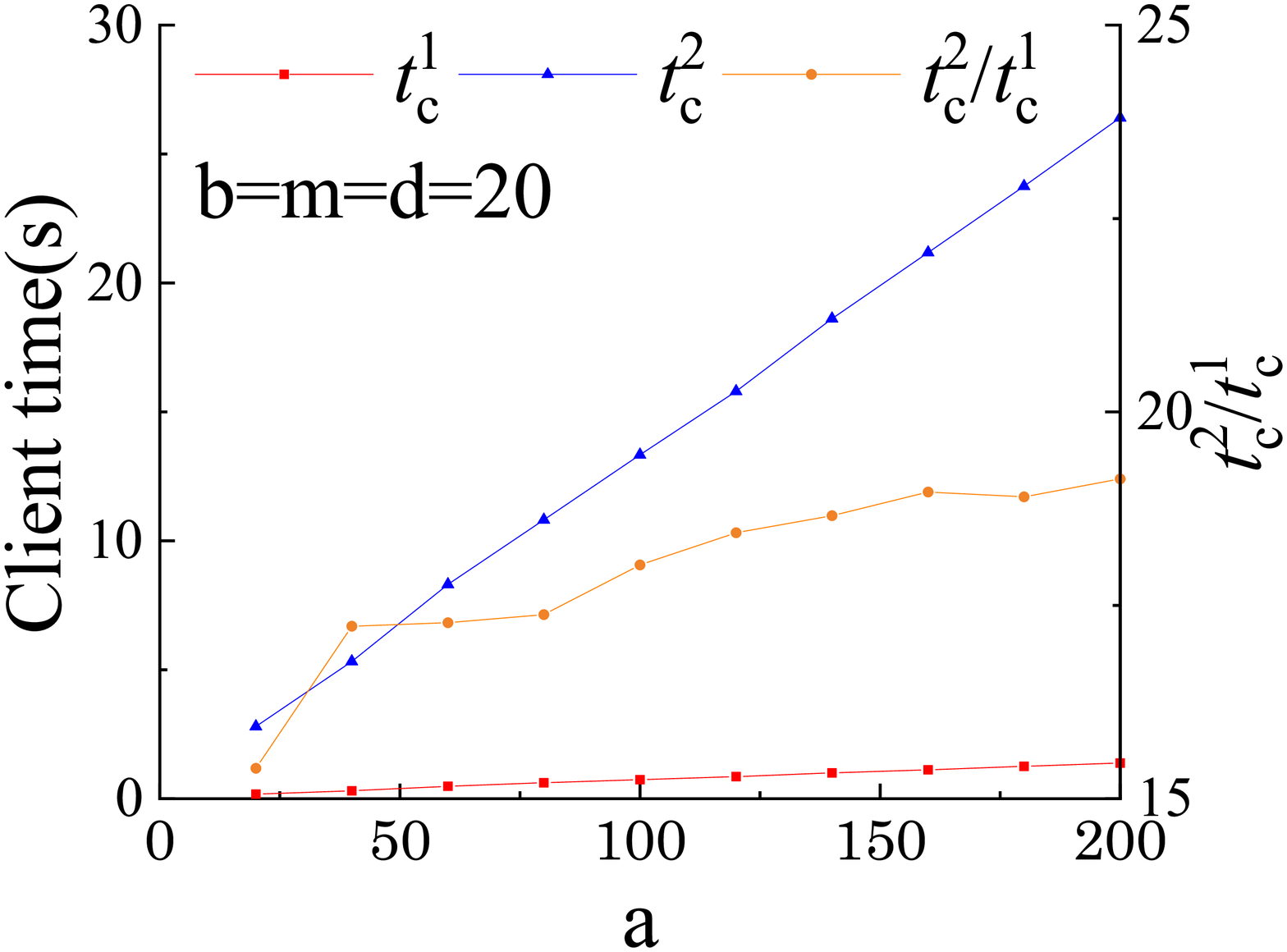}
}
\subfigure{
\includegraphics[width=0.23\textwidth]{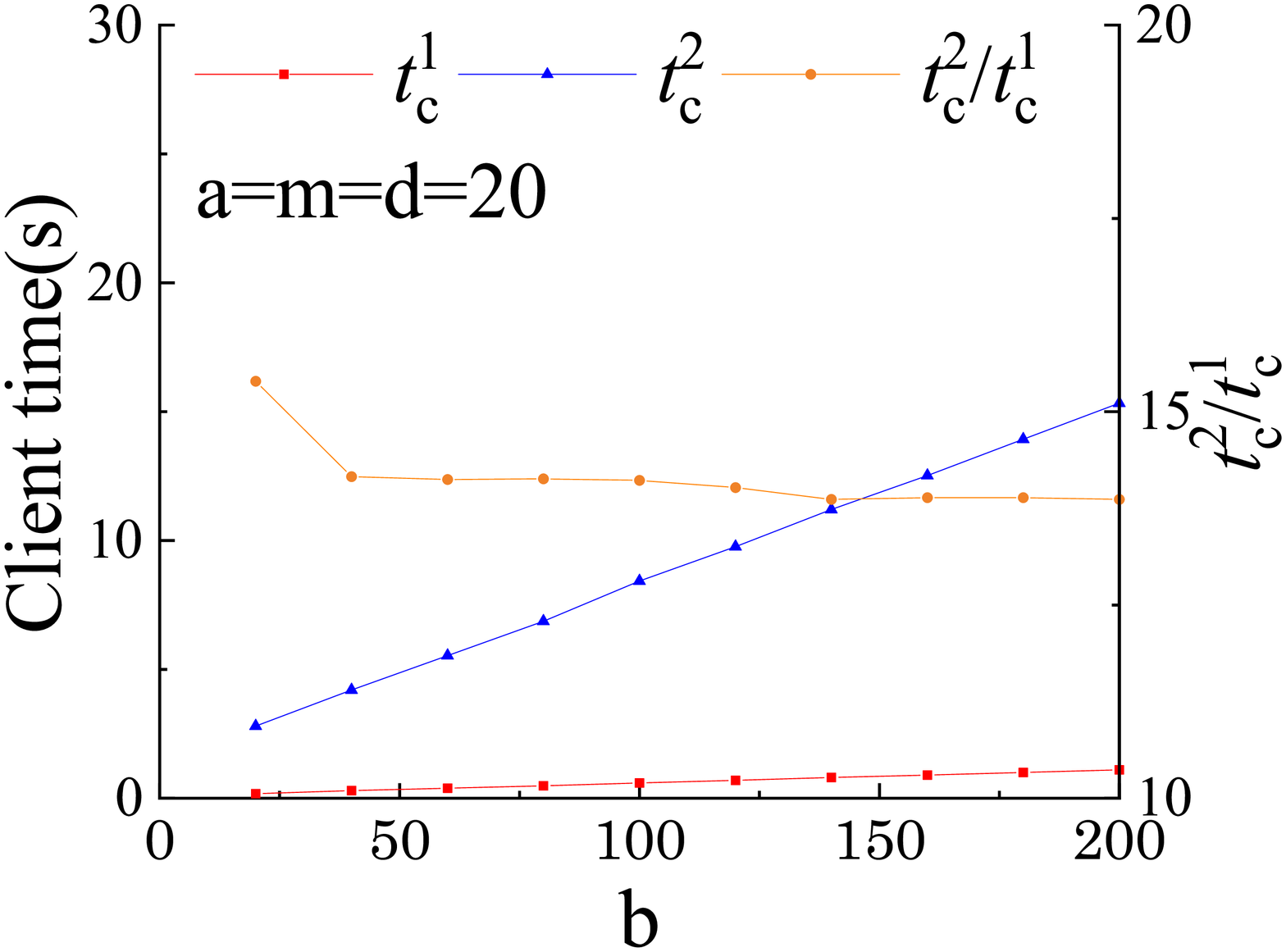}
}
\subfigure{
\includegraphics[width=0.23\textwidth]{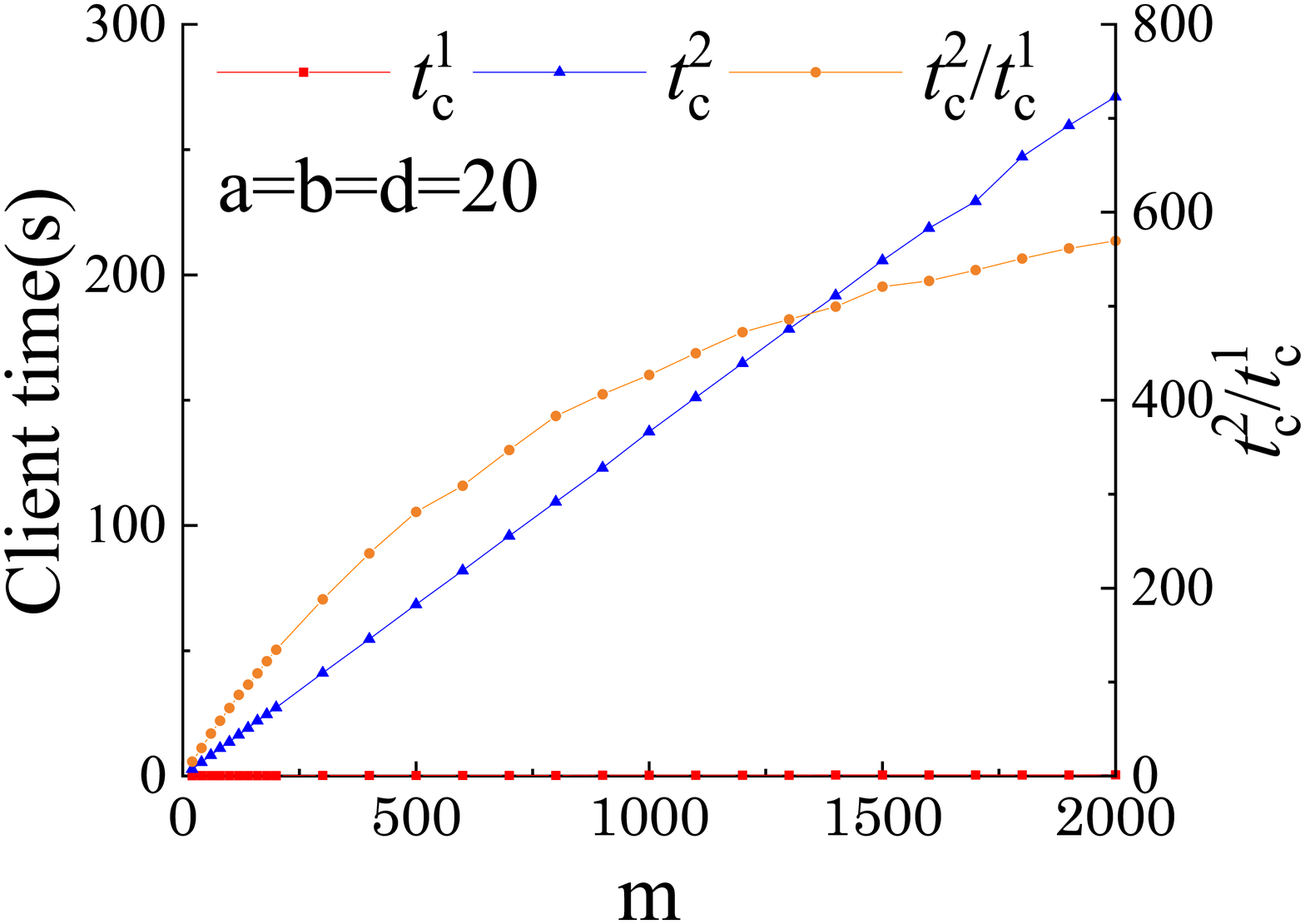}
}
\subfigure{
\includegraphics[width=0.23\textwidth]{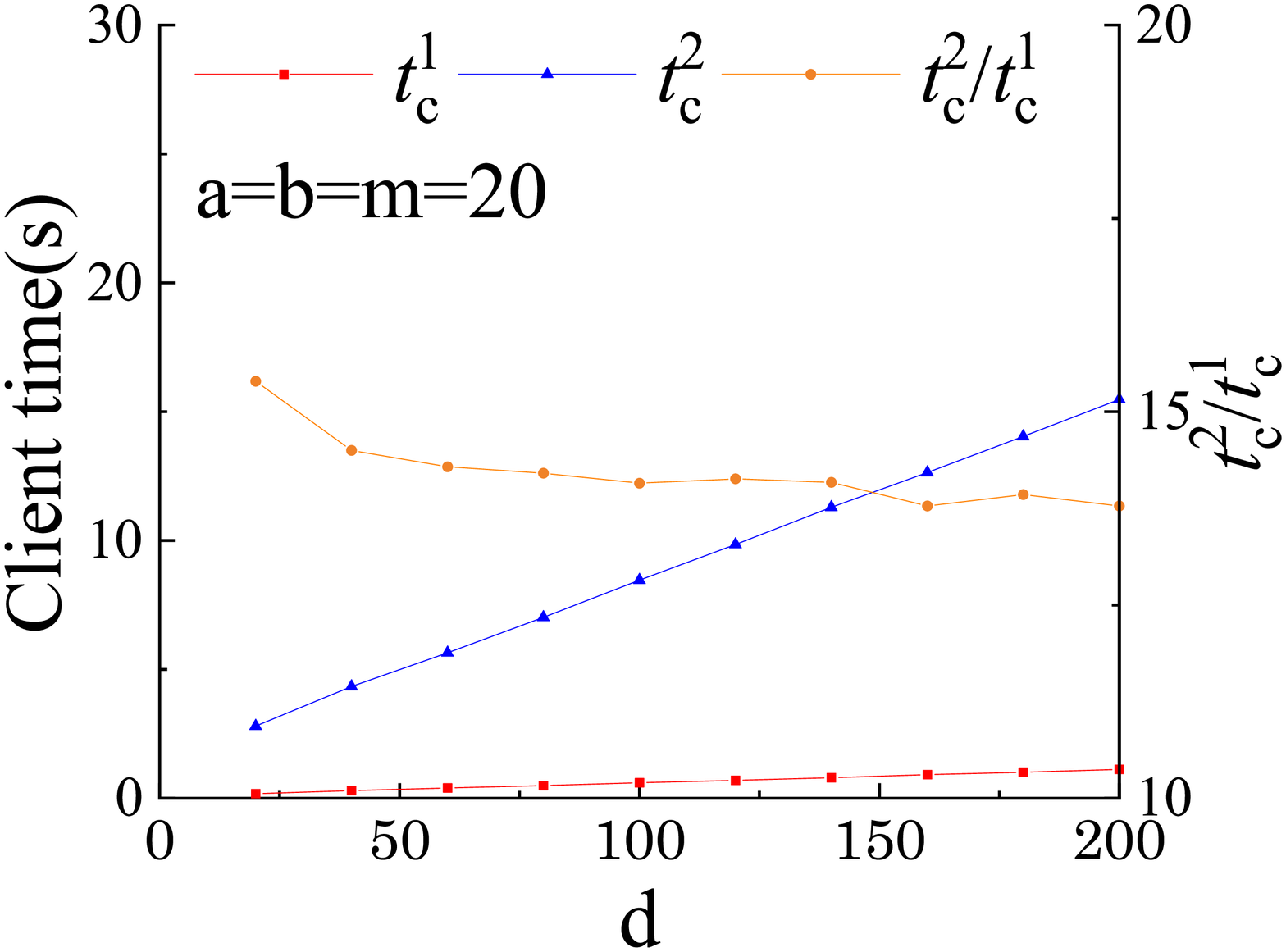}
}
\caption{\color{black} Client-side  computation time} \label{f1}
\end{figure*}
\begin{figure*}[htp]
\subfigure{
\includegraphics[width=0.23\textwidth]{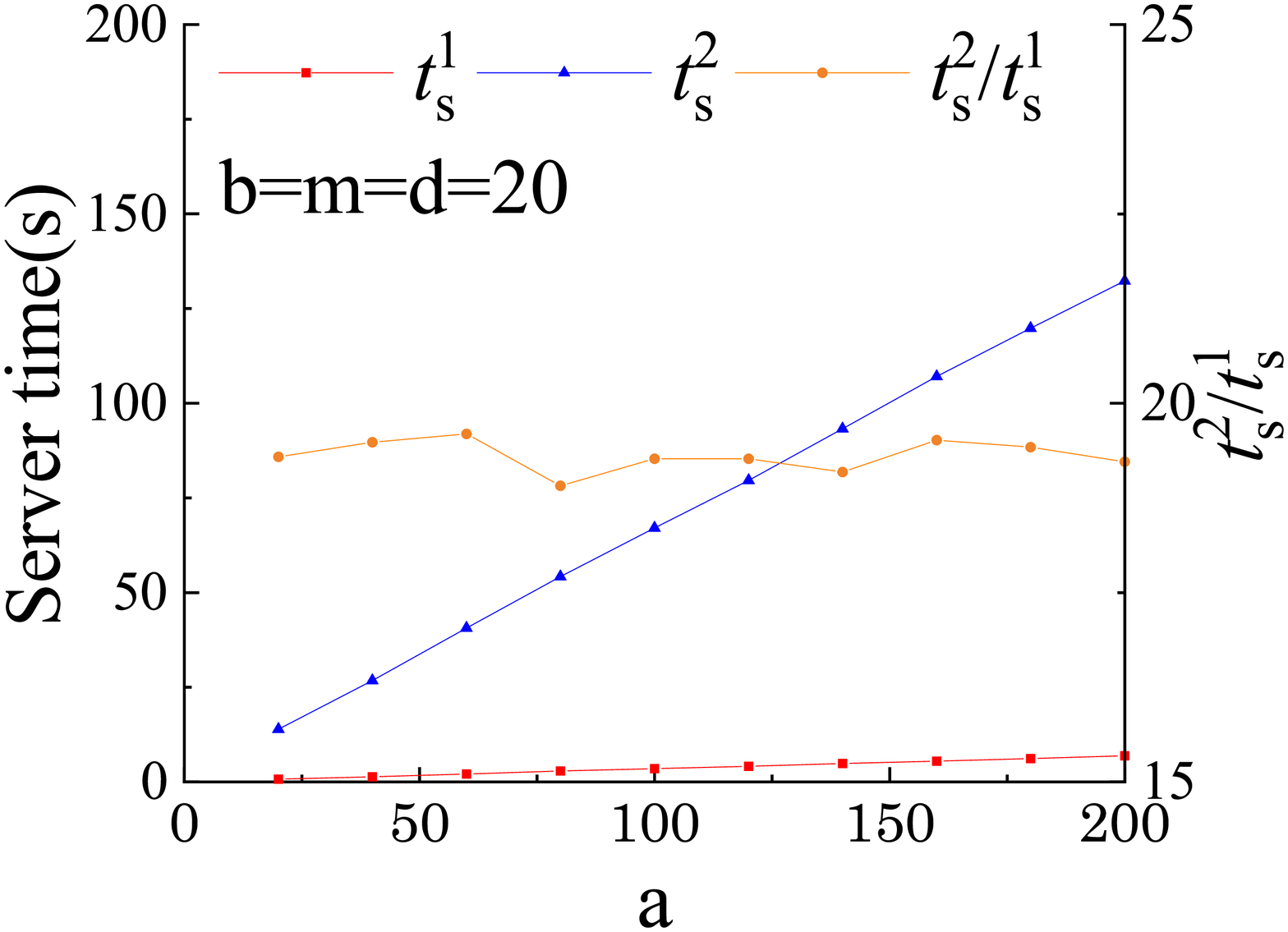}
}
\subfigure{
\includegraphics[width=0.23\textwidth]{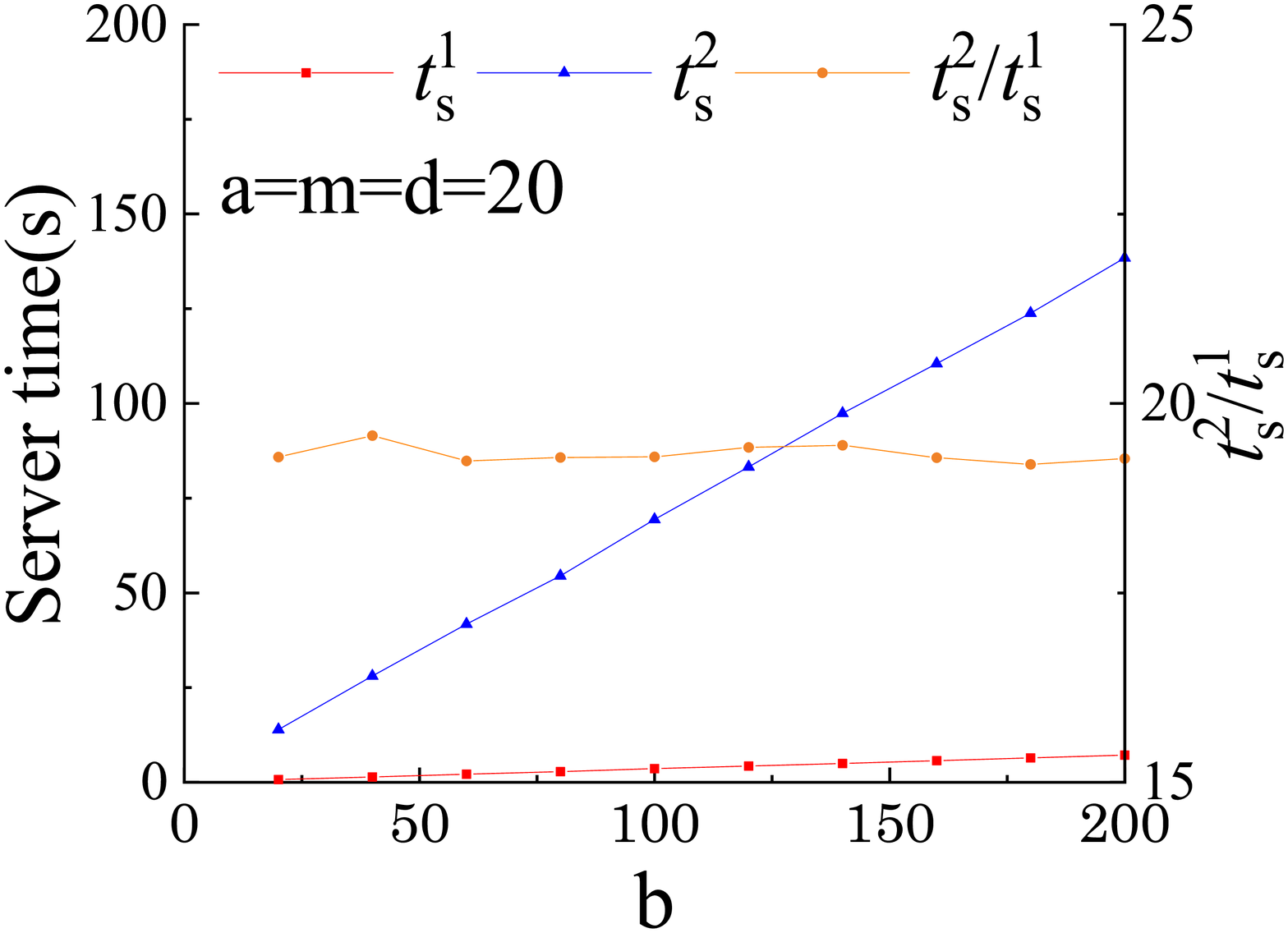}
}
\subfigure{
\includegraphics[width=0.23\textwidth]{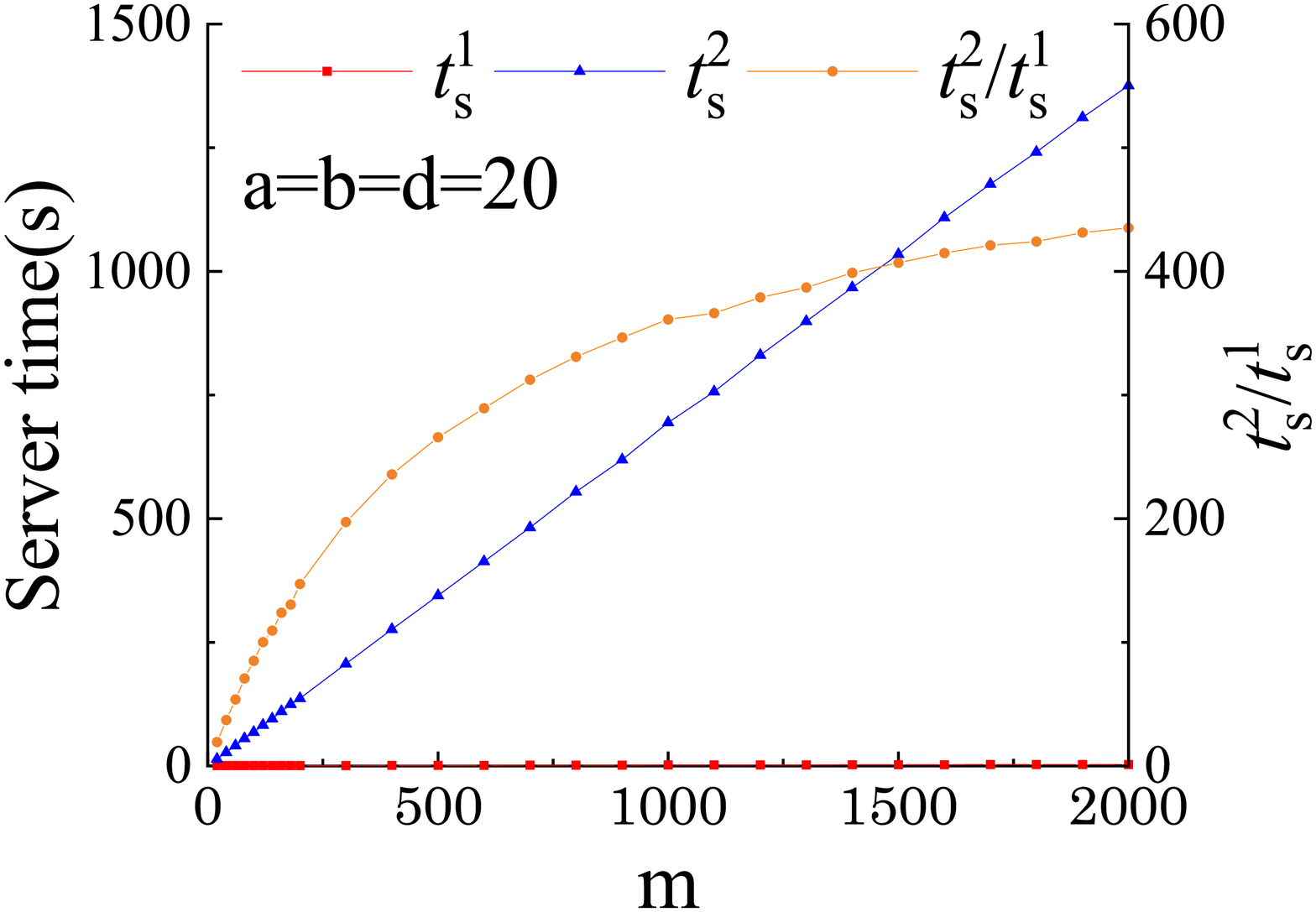}
}
\subfigure{
\includegraphics[width=0.23\textwidth]{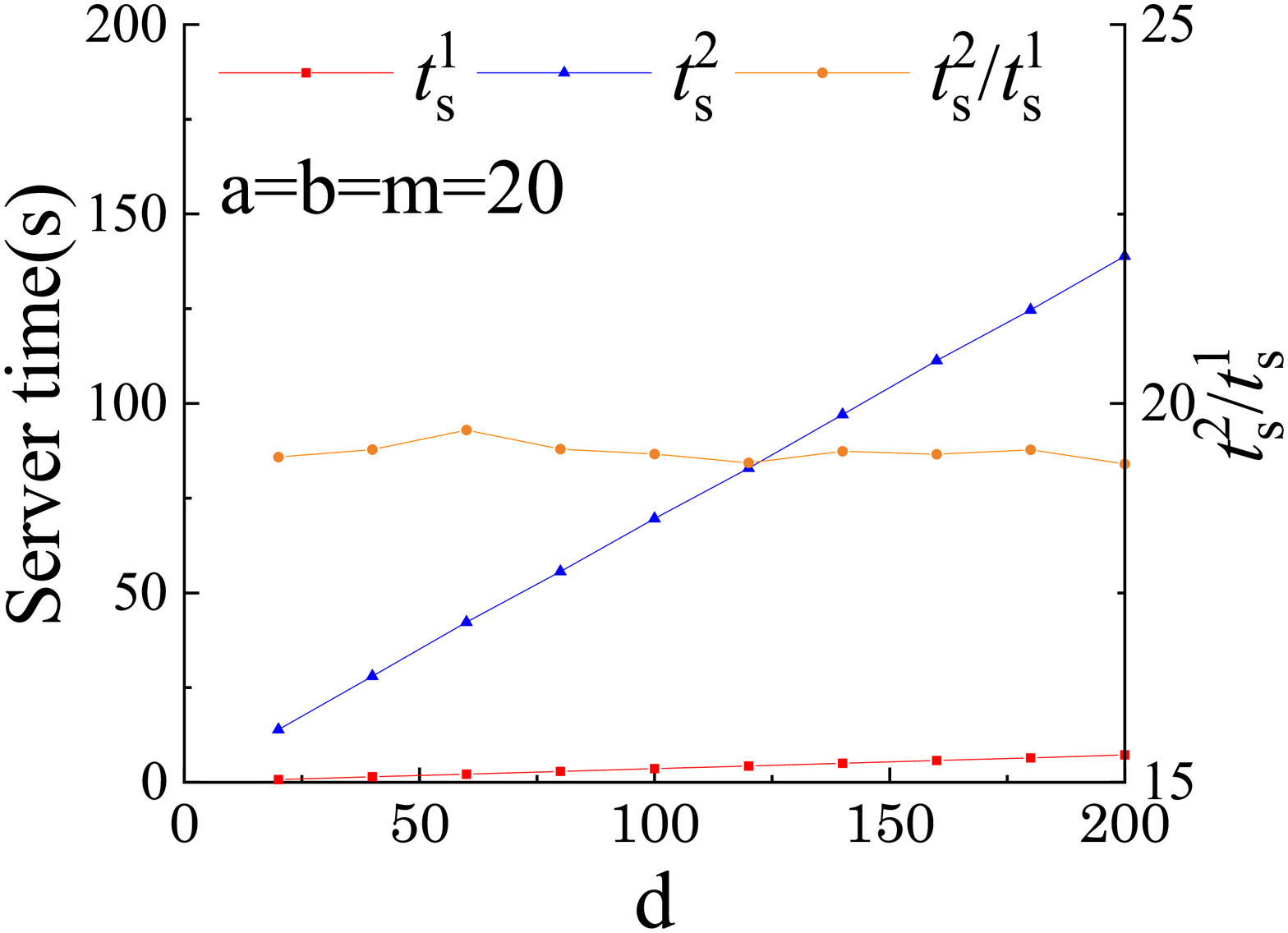}
}
\caption{\color{black} Sever-side  computation time} \label{f2}
\end{figure*}

\begin{figure*}
\subfigure{
\includegraphics[width=0.23\textwidth]{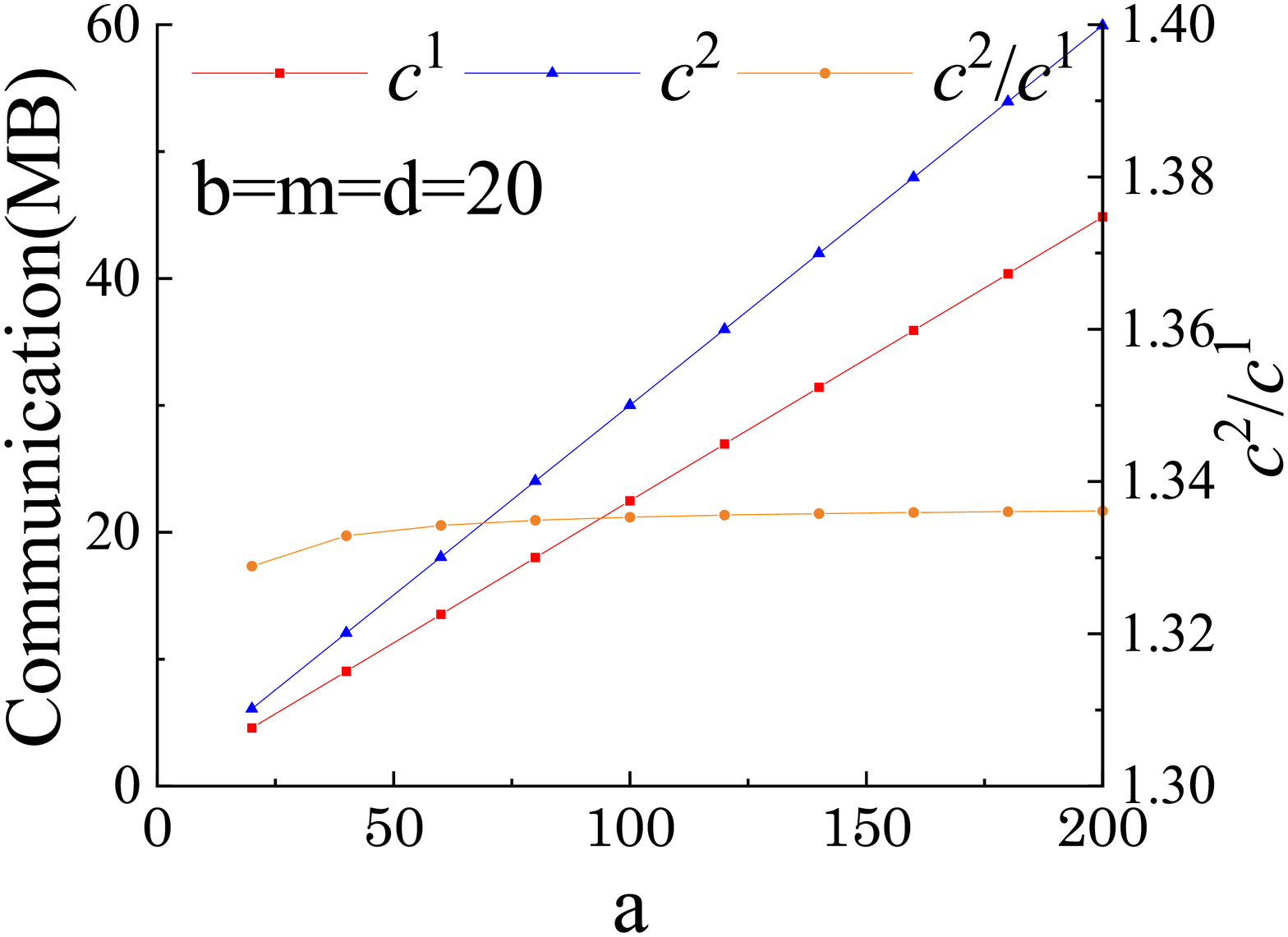}
}
\subfigure{
\includegraphics[width=0.23\textwidth]{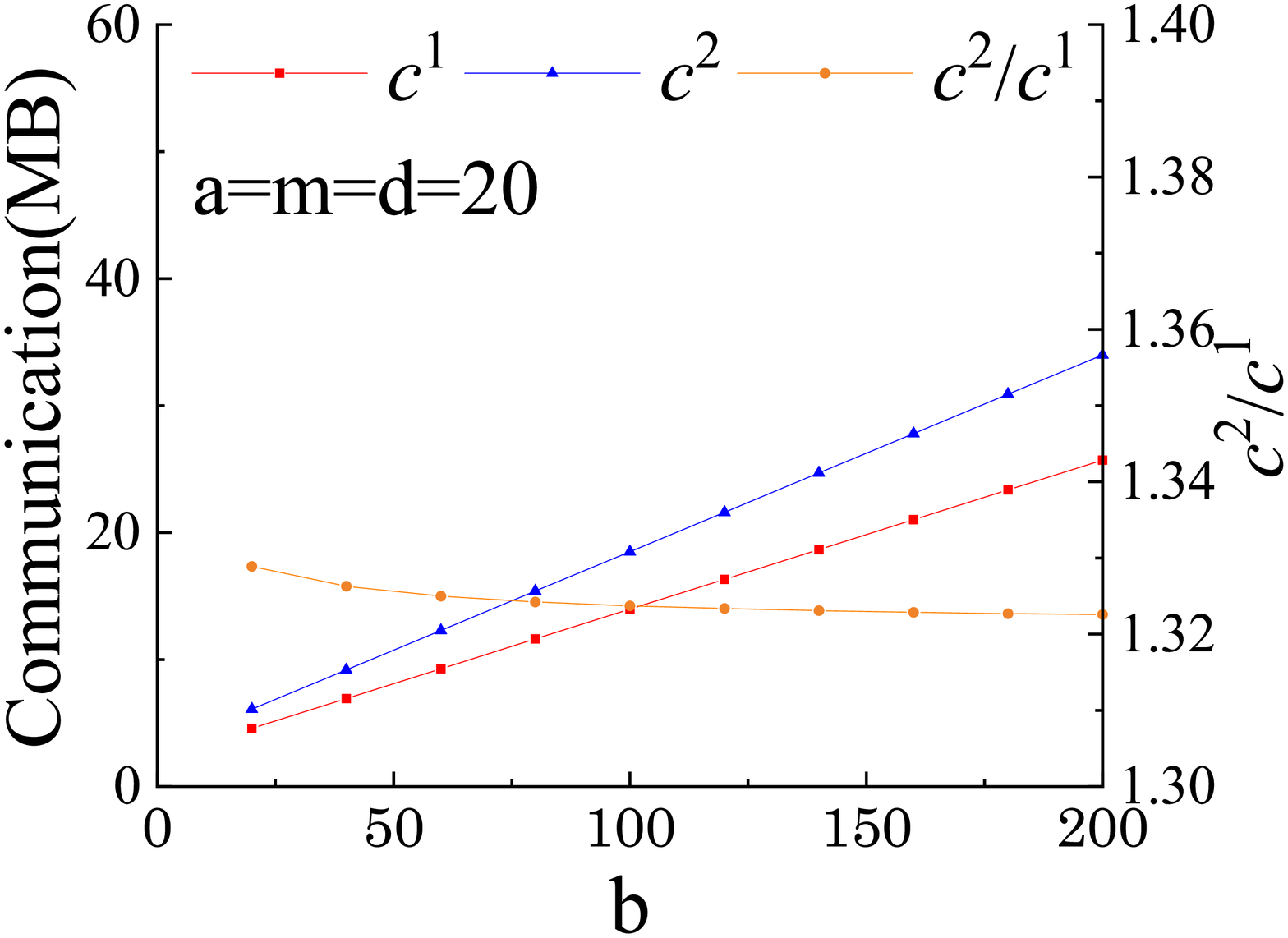}
}
\subfigure{
\includegraphics[width=0.23\textwidth]{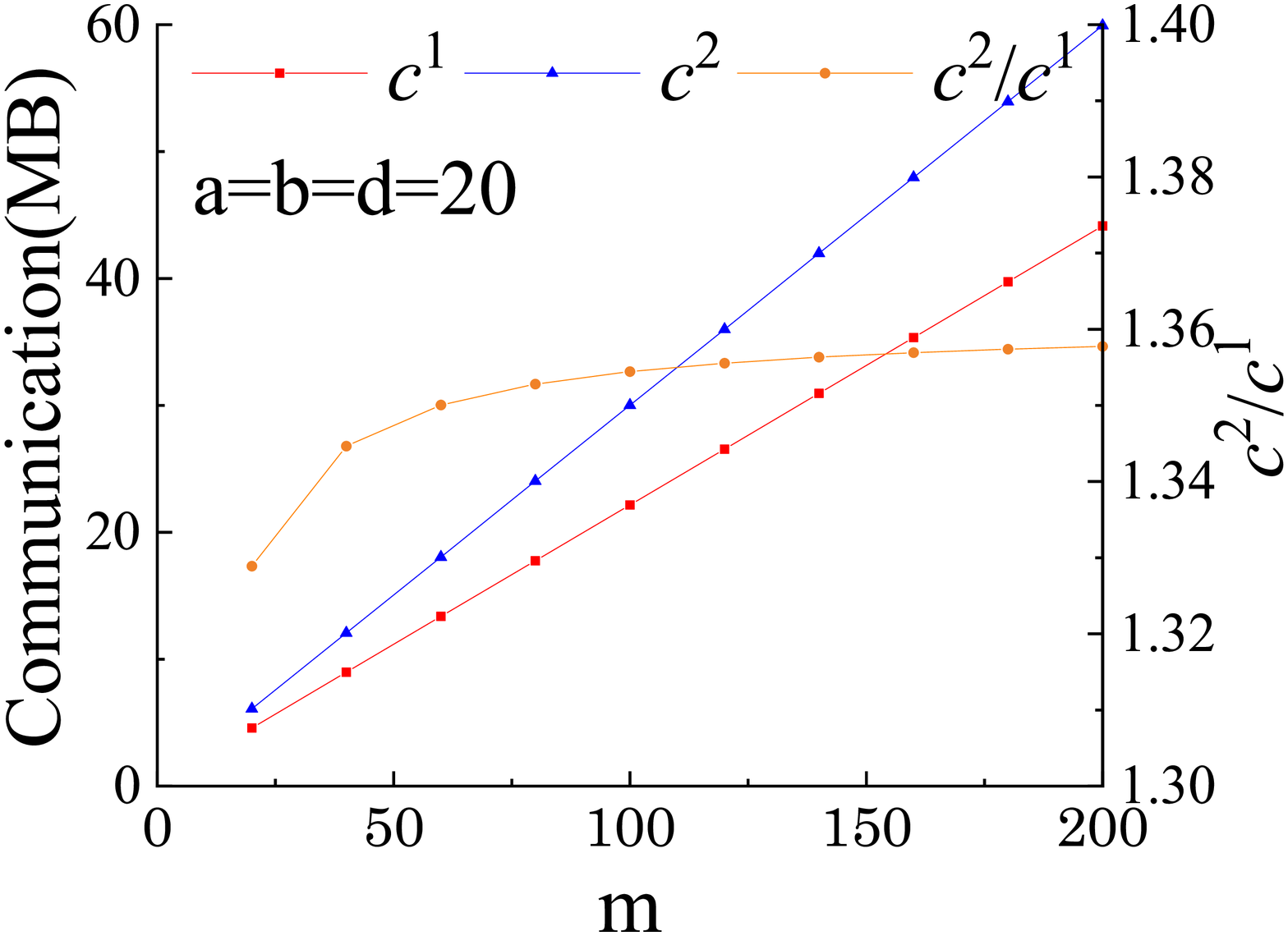}
}
\subfigure{
\includegraphics[width=0.23\textwidth]{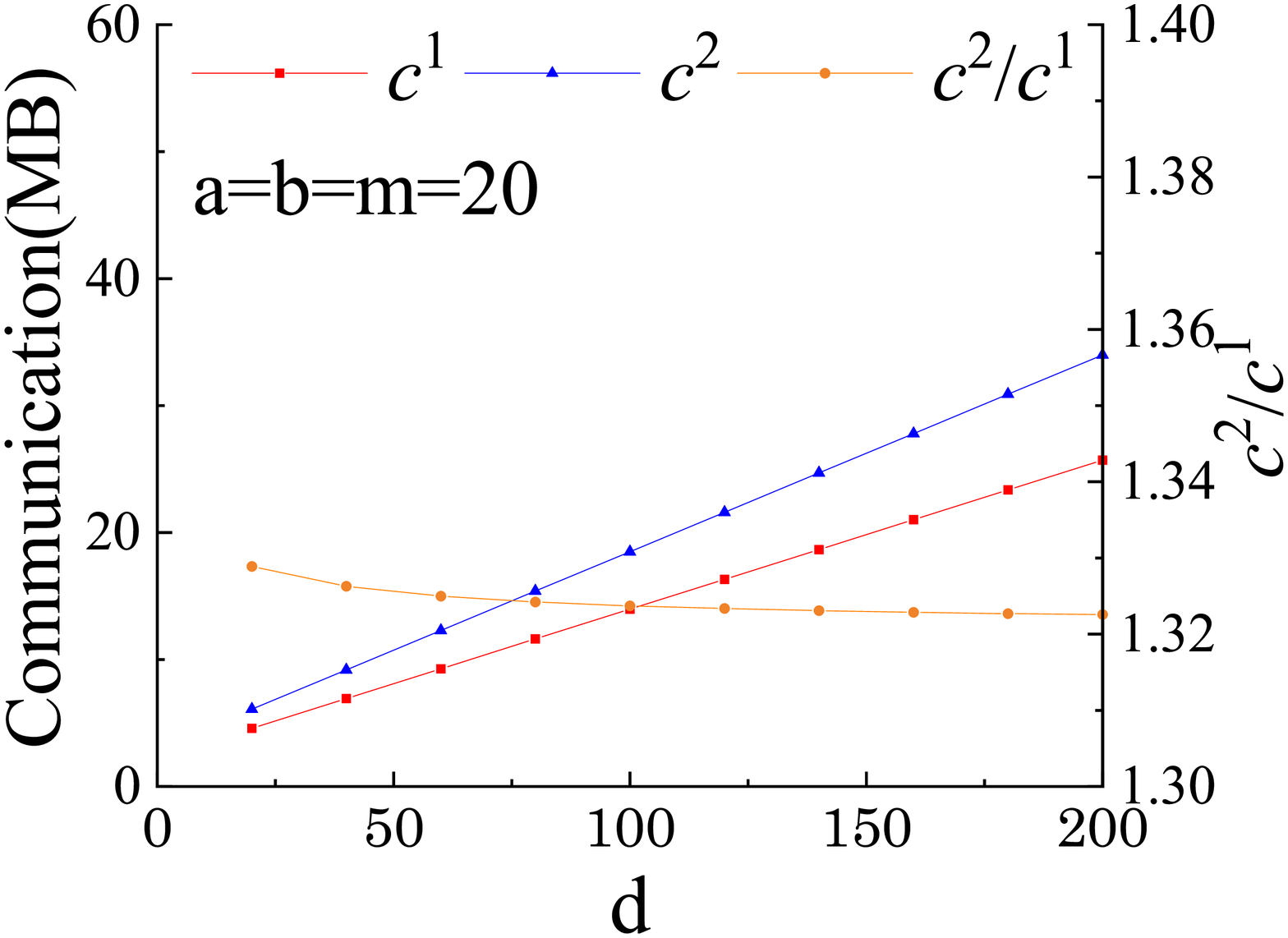}
}
\caption{\color{black} Communication complexity}  \label{f3}
\end{figure*}

\vspace{2mm}
\noindent
{\bf Storage complexity.}
 For every ${\bf F}\in \{{\bf F}_1,{\bf F}_2,
\ldots, {\bf F}_a\}$,
our scheme requires   the client to store two keys $VK_{\bf F}$  and $VK_{\bf x}$
for  future verification. In particular, $VK_{\bf F}=$
$(k,{\bf r})$
consists of  $m+1$ elements in $\mathbb{Z}_p$, and $VK_{\bf x}$ is an element of
$\mathbb{G}$.

\begin{threeparttable}[!htbp]
\center\caption{Storage Complexity}\label{Table3}
\small
\setlength{\tabcolsep}{5mm}{
\begin{tabular}{|c|c|c|}
\hline
&Elements in $\mathbb{Z}_p$&Elements in $\mathbb{G}$\\
\hline
\multirow{2}{*}{$VK_{\bf F}$}&$a(m+1)$&$0$\\
&\cellcolor[HTML]{C0C0C0}$2am$&\cellcolor[HTML]{C0C0C0}$0$\\
\hline
\multirow{2}{*}{$VK_{\bf x}$}&$0$&$b$\\
&\cellcolor[HTML]{C0C0C0}$0$&\cellcolor[HTML]{C0C0C0}$b$\\
\hline
\end{tabular}}
\begin{itemize}
\item non-shaded numbers: our storage complexity
\item shaded numbers: storage complexity of \cite{FG12}
\end{itemize}
\end{threeparttable}

\vspace{2mm}
\noindent
Table \ref{Table3} provides both a summary of the above analysis and  comparisons between our scheme
and   \cite{FG12} for  outsourcing the $ab$ computations $\{{\bf F}_i {\bf x}_j: i\in[a],j\in[b]\}$.
In particular, the non-shaded numbers   describe  our scheme and
the  shaded numbers  describe \cite{FG12}.
{\color{black}
We denote with
$s^1$ (resp. $s^2$) the storage complexity of
our scheme (resp. \cite{FG12}).  Then Table \ref{Table3}  shows that
\begin{equation*}
\begin{split}
s^1&=a(m+1)\ell_p+b\ell_{\mathbb{G}};\\
s^2&=2am\ell_p+b\ell_{\mathbb{G}}.
\end{split}
\end{equation*}
It's easy to see that $s^1<s^2$, i.e., the storage complexity of
 our scheme is always  smaller than  \cite{FG12}.
In particular,  when  $\ell_p=O(\ell_{\mathbb{G}})$
and  $am\gg b$, we will have
\begin{equation}
\label{eqn:stor}
s^2/s^1\approx 2.
\end{equation}
 }

\vspace{-1cm}

\subsection{Experimental Results}
\label{sec:exp}

We implemented  both our scheme and the scheme of \cite{FG12} for
outsourcing the computations of ${\bf Fx}$ for all
${\bf F}\in \{{\bf F}_1,{\bf F}_2,
\ldots, {\bf F}_a\}$ and
${\bf x}\in \{{\bf x}_1,{\bf x}_2,\ldots,{\bf x}_b\}$.
Our implementations are based on the RELIC toolkit in C language, and using OpenMP for threading support. All
executions are conducted on a computer with Intel(R)
Core(TM) i7-6700 CPU processor running at 3.40GHz and a
8GB RAM.

\begin{figure*}[htp]
\subfigure{
\includegraphics[width=0.23\textwidth]{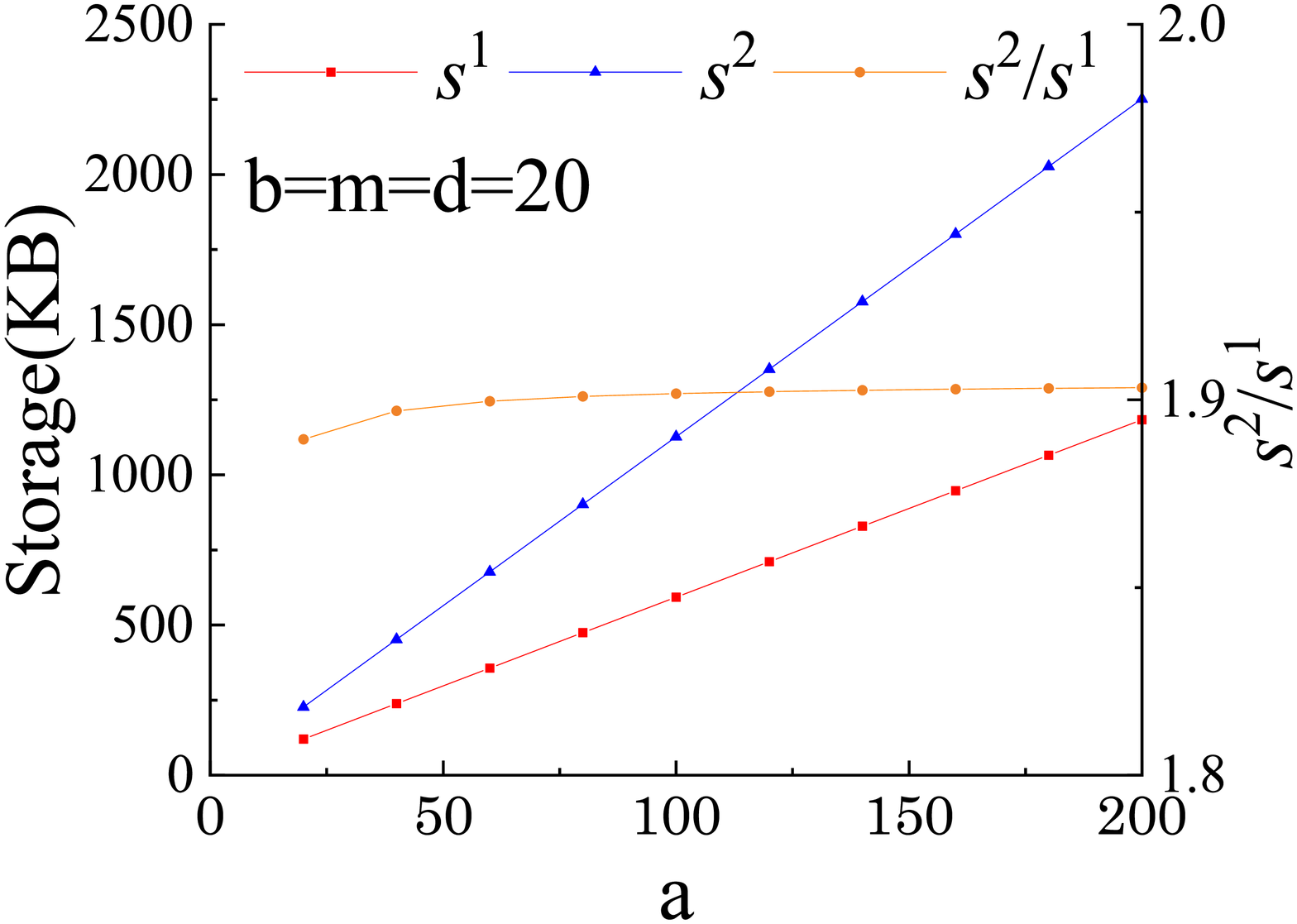}
}
\subfigure{
\includegraphics[width=0.23\textwidth]{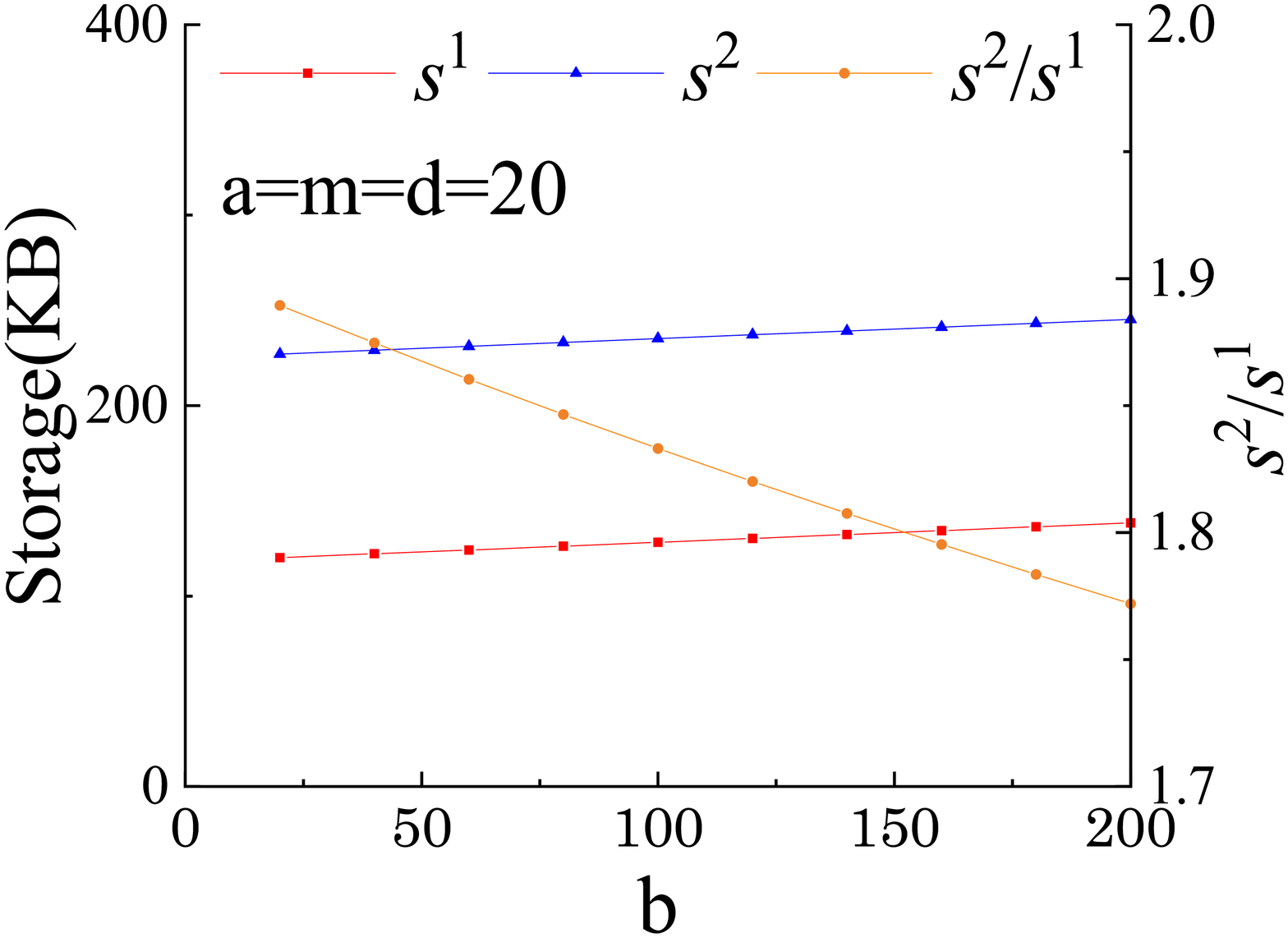}
}
\subfigure{
\includegraphics[width=0.23\textwidth]{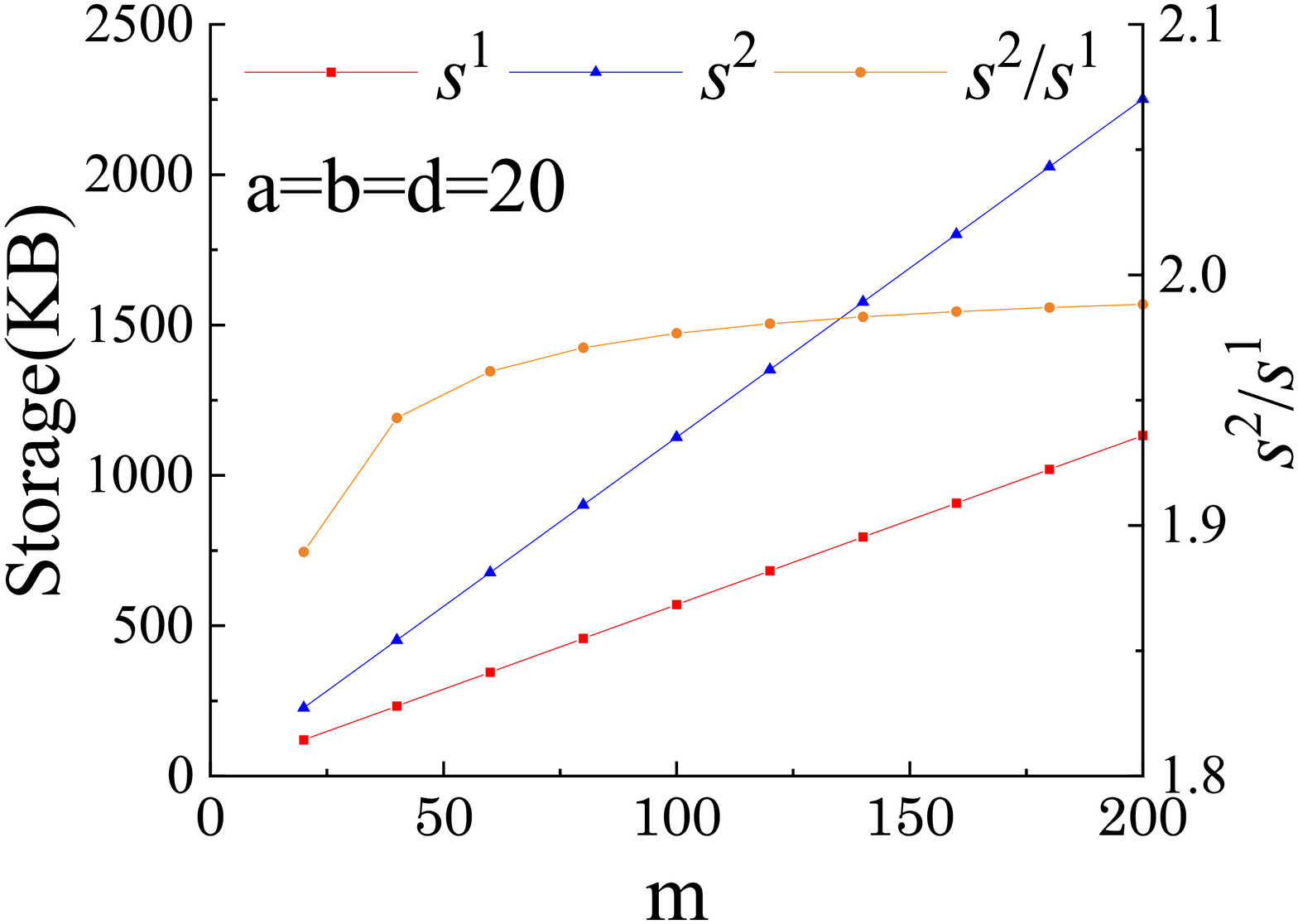}
}
\subfigure{
\includegraphics[width=0.23\textwidth]{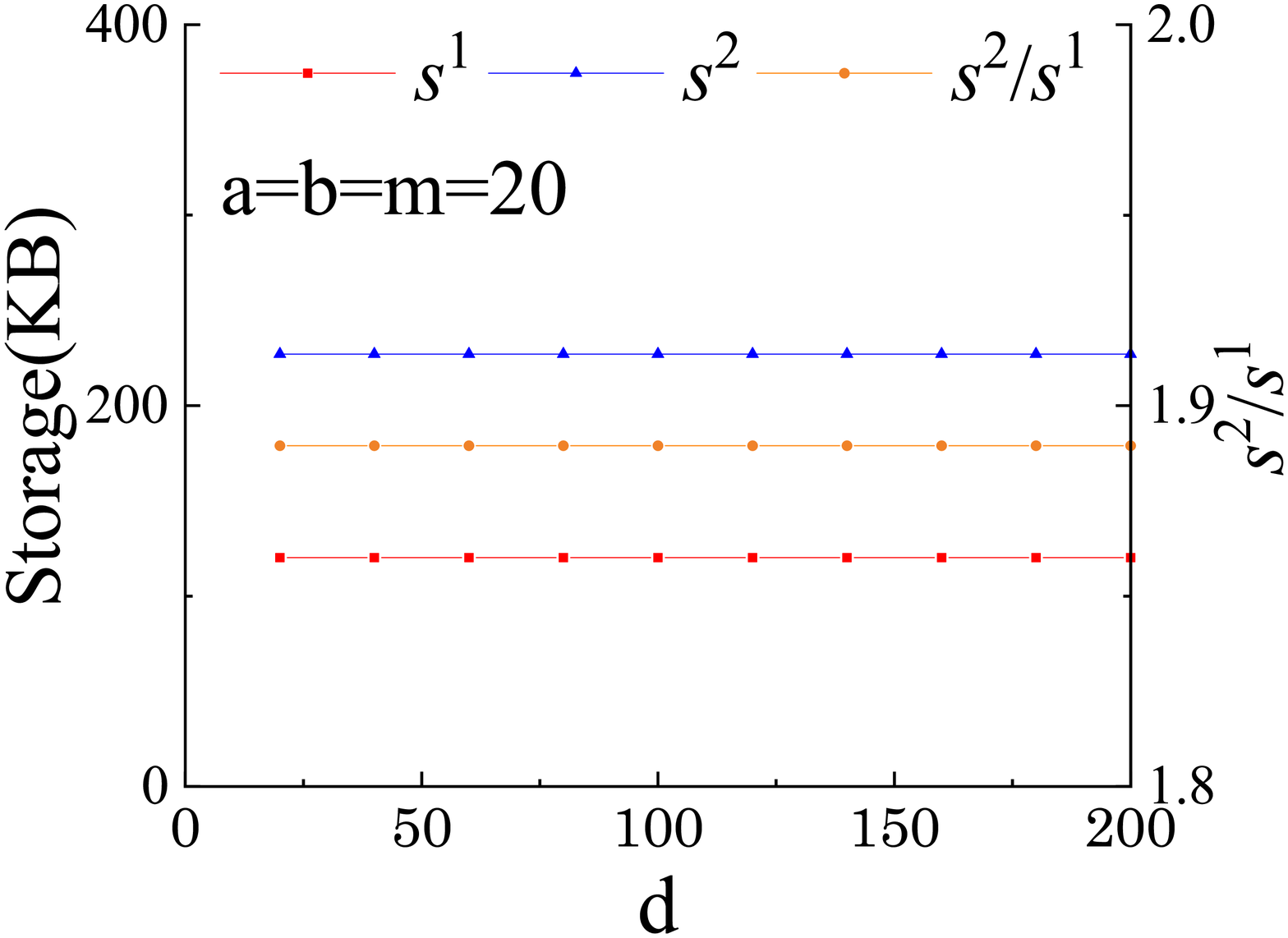}
}
\caption{\color{black} Storage complexity} \label{f4}
\end{figure*}

\vspace{2mm}
\noindent
{\bf Computation Complexity.}
 In our experiment,
we fix any three out of the four  parameters   $a,b,m$ and $d$, and
let the remaining parameter vary
in a certain range.
 Figure \ref{f1} shows the dependence of the client-side running time
 as a function of the remaining parameter.
 Figure \ref{f1} shows that
the client-side computation time in our scheme is always
  smaller than that of \cite{FG12}
   and the time saving  is consistent with the theoretical
  analysis below Table  \ref{Table1}.
{\color{black}  For example, when $a=b=m=20$ and $d=200$, our benchmark shows that
$t_{{\bf exp}_{\mathbb{G}}}\gg
\max\{t_{{\bf mul}_{\mathbb{G}}},mt_{{\bf mul}_p},
mt_{{\bf add}_p}\}$; our experiment shows that
  $t_{\rm c}^1\approx 1.12 {\rm s}, t_{\rm c}^2\approx 15.48{\rm s}$ and
  $ t_{\rm c}^2/ t_{\rm c}^1\approx 13.82 \geq 2m/3$, which is implied by Equation
  (\ref{eqn:time}).} Figure \ref{f2} shows that
  the server-side computation time in our scheme is always
  smaller than that of \cite{FG12}
   and the time saving  is consistent with the theoretical
  analysis below Table  \ref{Table1}.
{\color{black}  For example, when $a=b=m=20$ and $d=200$, our experiment shows that
  $t_{\rm s}^1\approx 7.23 {\rm ~s}, t_{\rm s}^2\approx 138.90{\rm ~s}$ and
  $ t_{\rm s}^2/ t_{\rm s}^1\approx 19.2$, which is  very close to $m$.
  This fact is also  implied by Equation
  (\ref{eqn:time}). }

\vspace{2mm}
\noindent
{\bf Communication Complexity.}
Figure \ref{f3} compares   the  communication complexity of
our scheme and \cite{FG12}.
  In our experiment, we choose the sets $\mathbb{Z}_p$ and $\mathbb{G}$
  such that each element of  $\mathbb{Z}_p$ has a representation of
   2304 bits and each element of $\mathbb{G}$ has a representation of
 832 bits, i.e., $\ell_p=2304$ and
  $\ell_{\mathbb{G}}=832$. Figure \ref{f3} shows that
   our  communication complexity is
  smaller than \cite{FG12}
   and the communication saving  is consistent with the theoretical
  analysis below Table  \ref{Table2}.
 {\color{black} For example, when $a=b=d=20$ and $m=200$, our experiment shows that
  $c^1\approx 44.13 {\rm MB}, c^2\approx 59.92{\rm MB}$ and $c^2/c^1\approx 1.36\approx
  1+\ell_{\mathbb{G}}/\ell_p$, which is implied by Equation
  (\ref{eqn:comm}). }

\vspace{2mm}
\noindent
{\bf Storage Complexity.}
Figure \ref{f4} compares  the  storage complexity of
our scheme and \cite{FG12}.
It shows that
   our  storage complexity is
  smaller than \cite{FG12}
   and the storage saving  is consistent with the theoretical
  analysis below Table  \ref{Table3}.
 {\color{black} For example, when $a=b=d=20$ and $m=200$, our experiment shows that
  $s^1\approx 1132.67 {\rm KB}, s^2\approx 2252.03{\rm KB}$ and $s^2/s^1\approx 2$,
  which is   implied by Equation
  (\ref{eqn:stor}). }

 }

\section{Conclusions}

In this paper, we constructed the first multi-function verifiable computation
 scheme for outsourcing matrix functions.
When it is used to outsource $m$ linear functions, the scheme outperforms
the scheme of \cite{FG12} by a factor of $m$.
This gives essential cost saving as long as
  $m$ grows and is large enough.
  Our technique of combining $m$ linear functions as one and then
  conduct a known verification may be of independent interest.
Our   multi-matrix verifiable computation scheme is publicly delegatable and private verifiable, it is an open
 problem  to construct a scheme that is both publicly delegatable and public verifiable.
As all previous multi-function verifiable computation schemes \cite{PRV12,FG12},
 ours does not protect the confidentiality of the client's  functions, inputs, or outputs.
It is also an interesting problem to construct a scheme that keeps the confidentiality of the client's data.

\vspace{2mm}
\noindent{\bf Acknowledgements.}  The authors would like to thank the anonymous
referees for their helpful comments. This work is supported by the
NSFC (No. 61602304).

\bibliographystyle{plain}

\end{document}